\documentclass{elsart}

\usepackage{amsmath}
\usepackage{amssymb}
\usepackage{latexsym}
\usepackage[dvips]{graphicx}

\journal{Astroparticle Physics}

\begin{document}

\begin{frontmatter}


\title{Neutrino Telescope Modelling of Lorentz Invariance Violation
in Oscillations of Atmospheric Neutrinos}

\author[Applied]{Dean Morgan}
\author[Applied]{Elizabeth Winstanley\corauthref{cor}}
\ead{E.Winstanley@sheffield.ac.uk}
\author[Marseille]{Jurgen Brunner}
\author[Physics]{Lee F. Thompson}

\address[Applied]{Department of Applied Mathematics,
The University of Sheffield, Hicks Building, Hounsfield Road,
Sheffield, S3 7RH, U.K.}
\address[Marseille]{Centre de Physique des Particules de Marseille,
163 Avenue de Luminy - \\ Case 907, 13288 Marseille Cedex 09, France.}
\address[Physics]{Department of Physics and Astronomy,
The University of Sheffield,\\ Hicks Building, Hounsfield Road, Sheffield, S3 7RH, U.K.}

\corauth[cor]{Corresponding author}

\begin{abstract}
One possible feature of quantum gravity may be the violation of Lorentz invariance.
In this paper we consider one particular manifestation of the violation of
Lorentz invariance, namely modified dispersion relations for massive neutrinos.
We show how such modified dispersion relations may affect atmospheric neutrino oscillations.
We then consider how neutrino telescopes, such as ANTARES, may be able to place bounds
on the magnitude of this type of Lorentz invariance violation.
\end{abstract}


\end{frontmatter}

\section{Introduction}
\label{sec:intro}

The search continues for a theory which describes gravity on quantum scales, and provides
us with
falsifiable experimental predictions (for a  review,
see, for example, \cite{Kiefer:2005uk}).
In recent years, the subject of quantum gravity phenomenology has rapidly grown
(see \cite{Amelino-Camelia:2004hm} for a review), complementing theoretical work.

One particular feature of quantum gravity which has attracted much attention in
particle and astro-particle physics phenomenology is the breaking of CPT symmetry
(see, for example, \cite{Mavromatos:2004sz} for a review).
CPT symmetry is anticipated to be broken in quantum gravity because the prerequisite
conditions of
quantum coherence, locality or Lorentz invariance may no longer
hold.
In previous work \cite{morgan:2004vv,Hooper:2004xr,Hooper:2005jp}
we have studied the effect of quantum decoherence on atmospheric and astrophysical
neutrino oscillations, and in the present work we turn our attention to
Lorentz invariance violation.
We focus on atmospheric neutrino oscillations, having considered Lorentz invariance
violating effects on astrophysical neutrinos in \cite{Hooper:2005jp}.
As in our previous work \cite{morgan:2004vv},
in this paper we take a phenomenological viewpoint.
We examine, independent of any one particular theory of quantum gravity,
what we may expect to observe if Lorentz symmetry is broken or deformed.

There is already a large body of work
relating to Lorentz invariance violation and neutrinos, some of which is briefly
reviewed in the following section.
Furthermore, existing atmospheric neutrino data has been thoroughly analyzed for
both Lorentz invariance violating effects and other new physics
(see, for example, \cite{Fogli:1999fs,Gonzalez-Garcia:2004wg,Giorgini:2005zd}).
Although the analysis of Super-Kamiokande and K2K data
\cite{Fogli:1999fs,Gonzalez-Garcia:2004wg}
strongly disfavours Lorentz invariance violation as either
the leading or sub-leading cause of atmospheric neutrino oscillations,
an analysis of high-energy atmospheric neutrino oscillations, as would
be observed at a neutrino telescope, would enable a wider range of Lorentz invariance
violating models to be probed, as well as offering the possibility of further strengthening
existing bounds on Lorentz invariance violating parameters.
This can be seen in the recent analysis of MACRO data \cite{Giorgini:2005zd}
using both low ($<$ 30 GeV) and high ($>$ 130 GeV) energy neutrinos.

In this paper, we consider the sensitivity of the ANTARES
neutrino telescope \cite{Korolkova:2004pg} to Lorentz invariance
violating effects although we expect a similar analysis to be applicable to other neutrino
telescopes,
such as ICECUBE \cite{Ahrens:2002dv}, NESTOR \cite{Resvanis:1993gx} and NEMO \cite{Sapienza:2006en}.
Whilst neutrino telescopes are primarily designed to observe neutrinos
from astrophysical and cosmological origins, atmospheric neutrinos form the
primary background to
these sources and it is likely that they will provide the first physics results.

The outline of this article is as follows. In section \ref{sec:theory},
we briefly discuss some of the theoretical
consequences from Lorentz invariance violation,
in particular modified dispersion relations.
We then model how
modified dispersion relations alter
atmospheric neutrino oscillations in section \ref{sec:mdr}.
We present the results of our
simulations for the ANTARES neutrino telescope in section \ref{sec:sim},
where we first consider how
the observed spectra of events may be altered by Lorentz invariance violating
effects and then present the results of our
sensitivity simulations.
In section \ref{sec:conc}, we
present our conclusions and compare the results obtained
here with those from previous analyses of experimental data
\cite{Fogli:1999fs,Gonzalez-Garcia:2004wg,Giorgini:2005zd}.
Unless otherwise stated, we use units in which $c=\hbar=1$.

\section{Quantum gravity violations of Lorentz invariance
and the effect on neutrino oscillations}
\label{sec:theory}


In quantum gravity, space-time is expected to take on a
foamy nature \cite{Wheeler:1}.
Models of space-time foam have been built in many frameworks,
including critical
\cite{Horowitz:2004rn,Mohaupt:2002py} and non-critical
string theory
\cite{Dhar:2001a},
and loop quantum gravity \cite{Rovelli:1997yv}.
Interactions with this
space-time foam may lead to the breaking of CPT symmetry, leading
to the violation of Lorentz invariance \cite{Greenberg:2002uu}.
In addition, some theories of quantum gravity predict
that there is a minimum length scale, of order the Planck length ($10^{-35}$ m)
\cite{Horowitz:2004rn,Rovelli:1994ge}.
If this is the case, then the existence of a fundamental length scale may also
lead to Lorentz invariance violation (LV).

To illustrate this, imagine if we were able to somehow measure this fundamental length scale.
We could measure it in our lab frame and probably find that it was of order
the Planck length, $10^{-35}$ m.
If we now repeated the experiment
in a different inertial frame
(moving relative to the first frame), we would expect the length to be
shorter, as predicted by the usual Lorentz contraction.
However, we
are considering a fundamental length scale and, therefore, we
would also expect to measure the same length in all inertial frames.
There is, therefore, either an error in our naive thought experiment or
Lorentz symmetry, at least at the Planck scale, is broken.

Therefore, there exist three possibilities for the fate of Lorentz
invariance within theories of quantum gravity:
\begin{enumerate}
\item
It may be that our naive thought experiment is wrong, or
that a fundamental length scale does not exist, and so Lorentz
invariance remains intact at Planck scales;
\item
It may be that, at the Planck scale, Lorentz invariance is
broken and there is a preferred frame;
\item
It may be that Lorentz invariance as we know it is deformed,
so, for example, the Lorentz transformation involves a second
quantity which is not observer dependent. So, perhaps the Planck
length is invariant in an identical way to the speed of light.
\end{enumerate}
In the second and third options, one consequence of such
breaking/deforming of Lorentz invariance may be that quantities we
think of as being Lorentz invariant, such as the standard
dispersion relation, $E^{2}=p^{2}+m^{2}$, may also have to be modified.

Lorentz invariance violation may manifest itself in many different ways
and in many different experimental systems \cite{Mattingly:2005re,Coleman:1998ti}.
Our focus here is on neutrinos and modified dispersion relations.
However, modifications of dispersion relations have also been considered for photons
\cite{Gambini:1998it,Ellis:1999yd}, leading to dispersion relations of
the form \cite{Ellis:1999yd}:
\begin{displaymath}
\omega = k\left(1-\frac{k}{M}+O(M^{-2})\right),
\end{displaymath}
where $\omega$ is the frequency of the photon, $k$ its momentum and $M$ is the mass
scale of the quantum gravitational theory.
By considering photons which originate in
astrophysical objects such as gamma ray bursts and pulsars,
lower bounds of $\sim 10^{17}$ GeV \cite{Ellis:1999yd}
were placed on $M$ by considering  the delay in arrival times.

In the literature, the effects of LV on neutrino oscillations can be broken down into two
distinct categories.
In the first category are those models in which the only modification is to the
neutrino oscillation length
\cite{Hooper:2005jp,Lambiase:2003bq,Christian:2004xb,kelley:2006},
which is the type of LV model we shall consider in this paper.

In the second category are those models in which both the oscillation length and also
the mixing angle are altered.
In our simulations, we found that including LV effects on the
mixing angle as well as the neutrino oscillation length made no difference to our
results and so we do not study this type of model in detail.
This type of LV, where both the mixing angle and oscillation length are altered, has been considered
for the ICECUBE \cite{Gonzalez-Garcia:2006mh} and MINOS \cite{Blennow:2007pu} experiments in addition to
future neutrino factories \cite{Blennow:2005qj}.

\section{Modified dispersion relations and neutrino oscillation \newline probabilities}
\label{sec:mdr}

Our interest in this paper is how LV effects, as described in the previous
section, may alter the oscillations of atmospheric neutrinos. Therefore,
in the analysis which follows, we restrict ourselves to a simple
two neutrino system although it is possible to
consider how these effects manifest themselves in a
three neutrino system \cite{Hooper:2005jp}.

If we allow LV then this may lead to
modifications of the dispersion relation
\cite{Lambiase:2003bq,Christian:2004xb,Amelino-Camelia:2002fw,Amelino-Camelia:2003ex}:
\begin{equation}
\label{lvtheory:eqn:moddisp}
E^{2}=p^{2}+m^{2}+f(p,E,E_{p}),
\end{equation}
where $E$ is the energy of the neutrino, $p$ the neutrino
momentum, $m$ is the mass eigenvalue of the neutrino and $E_{p}$
is the Planck energy.
The function $f$ contains all the novel LV effects.
For our analysis here, we find it useful to parameterize
this function so the dispersion relation
(\ref{lvtheory:eqn:moddisp}) becomes \cite{Amelino-Camelia:2003ex}
\begin{equation}
\label{lvtheory:eqn:parmoddisp}
E^{2}=p^{2}+m^{2}+\eta
p^{2}\left(\frac{E}{E_{p}}\right)^{\alpha},
\end{equation}
where $\eta$ and $\alpha$ are parameters of the Lorentz invariance
violating theory.

Using (\ref{lvtheory:eqn:parmoddisp}) and assuming the parameter
$\eta$ has a dependence upon the mass eigenstate, then we find
that the Hamiltonian in the mass basis, for a two neutrino system,
may be written, to first order in $\eta$, as
\begin{equation}
\label{lvtheory:eqn:2neutLVHam}
H=\left(%
\begin{array}{cc}
  \frac{m_{1}^{2}}{2E}+\frac{\eta_{1}{E^{\alpha+1}}}{2} & 0 \\
  0 & \frac{m_{2}^{2}}{2E}+\frac{\eta_{2}{E^{\alpha+1}}}{2} \\
\end{array}%
\right),
\end{equation}
where $m_{i}$ and $\eta_{i}$ are the mass and  LV parameter of the $i$th eigenstate.
Here we are considering only the case when the Hamiltonian is diagonal in the mass
basis, as off-diagonal terms will modify the neutrino mixing angle as well as the
oscillation length.

The probability that a neutrino of flavour $a$ oscillates to one
of flavour $b$ is then given by
\begin{displaymath}
P[\nu_{a}\rightarrow\nu_{b}] =
\left|\sum_{i=1}^{2}U_{ai}e^{-H_{ii}}U_{bi}^{*}\right|^{2},
\end{displaymath}
where the $U$'s are the components of the neutrino mixing matrix:
\begin{displaymath}
U=\left(%
\begin{array}{cc}
  \cos\theta & \sin\theta \\
  -\sin\theta & \cos\theta \\
\end{array}%
\right) ,
\end{displaymath}
with $\theta$ being the mixing angle.
The oscillation probability is therefore
\begin{displaymath}
P[\nu_{a}\rightarrow\nu_{b}] =
\sin^{2}2\theta\sin^{2}\left[\frac{\Delta
m^{2}L}{4E}+\frac{\Delta\eta
E^{\alpha+1}L}{4E_{p}^{\alpha}}\right],
\end{displaymath}
where $\Delta m^{2}=m_{2}^{2}-m_{1}^{2}$ and $\Delta \eta=\eta_{2}-\eta_{1}$.
Replacing $c$ and $\hbar$ and absorbing the Planck energy into the
parameter $\Delta\eta$ gives
\begin{equation}
\label{lvtheory:eqn:2nprobch}
P[\nu_{a}\rightarrow\nu_{b}] =
\sin^{2}2\theta\sin^{2}\left[1.27\frac{\Delta
m^{2}L}{E}+1.27\times10^{9(n+1)}\Delta\eta E^{n}L\right],
\end{equation}
where $\Delta m^{2}$ is measured in eV${}^{2}$, the neutrino energy $E$ in GeV,
the path length $L$ in km,
and we have set $n=\alpha+1$. If we set $\Delta\eta=0$ in
(\ref{lvtheory:eqn:2nprobch}), then we
recover the standard oscillation probability
\begin{equation}
\label{lvtheory:eqn:standprob}
P[\nu_{a}\rightarrow\nu_{b}] =
\sin^{2}2\theta\sin^{2}\left[1.27\frac{\Delta
m^{2}L}{E}\right],
\end{equation}
whereas, if we set $\Delta m^{2}=0$,
we find a probability in which neutrinos oscillate due to LV
effects only:
\begin{displaymath}
P[\nu_{a}\rightarrow\nu_{b}] =
\sin^{2}2\theta\sin^{2}\left[1.27\times10^{9(n+1)}\Delta\eta E^{n}L\right] .
\end{displaymath}
In (\ref{lvtheory:eqn:2neutLVHam}), we have assumed that the LV
parameter $\eta$ has a dependence upon the mass eigenstate. If
this is not the case and the parameter $\eta$ is universal to all
mass eigenstates, then, even if the theory violates Lorentz
invariance, the neutrino oscillation probability is unchanged.

In this paper, we shall consider three models with differing energy dependences, namely:
\begin{enumerate}
\item
Model LV1 - with $n=1$: here the LV parameter $\Delta \eta $
in (\ref{lvtheory:eqn:2nprobch}) is dimensionless;
\item
Model LV2 - with $n=2$: then $\Delta \eta $ has units of eV${}^{-1}$;
\item
Model LV3 - with $n=3$: in which case $\Delta \eta $ has units eV${}^{-2}$.
\end{enumerate}
It may be, in a phenomenology
of quantum gravity, that $n$ is not an integer but we only consider integer values here
(the results
for non-integer $n$ can be found by interpolating between the two bounding integer values).
Models with $n=-1$, $0$ and $1$ have been examined using Super-Kamiokande \cite{Fogli:1999fs}
data, and more complicated models with $n=0$ or $1$ have been analyzed using both
Super-Kamiokande and K2K data \cite{Gonzalez-Garcia:2004wg}.
We do not consider $n<1$ here as $n=0$ models have been severely constrained by these
previous analyses, and, since in that case the LV modifications of the oscillation probability
are independent of the neutrino energy, we do not expect to be able to improve on their
bounds by using higher-energy atmospheric neutrinos.
Therefore we only consider values of $n$ which are strictly positive, and include $n=1$
for comparison with previous work \cite{Fogli:1999fs,Gonzalez-Garcia:2004wg}.
From our work on quantum decoherence effects on atmospheric neutrino oscillations
\cite{morgan:2004vv}, one might anticipate that we will be able to place more stringent bounds
on LV effects as $n$ increases.

In the next section, we shall be concerned with finding upper
bounds on the LV parameters.
However, we are able to extract crude order of magnitude estimates for
these parameters from the oscillation probability
(\ref{lvtheory:eqn:2nprobch}).
The LV effects only become significant when
\begin{displaymath}
1.27\frac{\Delta m^{2}}{E}\sim 1.27\times10^{9(n+1)}\Delta\eta E^{n}.
\end{displaymath}
The peak in the atmospheric flux is of the order of $1$ GeV, which is the typical energy
scale for experiments such as Super-Kamiokande \cite{Ashie:2004mr},
whilst neutrino telescopes, such as ANTARES,
are sensitive to much higher energy neutrinos.
Table \ref{tab:estbounds} shows a comparison
of the typical bounds on $\Delta\eta$ for low and high energy neutrino experiments.
Here, we have taken the
peak in the atmospheric flux, $E\sim~1$ GeV as the value of the
neutrino energy for low energy experiments and $E\sim100$ GeV
as a typical value for neutrinos to be detected by a neutrino telescope such as ANTARES.
\begin{table}
\center{\begin{tabular}{|c|c|c|} \hline
 & $\Delta \eta $ & $\Delta \eta $ \\ \hline
$n$  & $E=1$ GeV  & $E=100$ GeV\\
\hline
1 & $2.3\times10^{-21}$ & $2.3\times10^{-25}$  \\
\hline 2 & $2.3\times10^{-30}$ eV${}^{-1}$ & $2.3\times10^{-36}$ eV${}^{-1}$ \\
\hline 3 & $2.3\times10^{-39}$ eV${}^{-2}$ & $2.3\times10^{-47}$ eV${}^{-2}$ \\
\hline
\end{tabular}}
\caption{Table showing the expected upper bounds on $\Delta\eta$,
the LV parameter for various values of $n$,
for atmospheric neutrinos with low energies, $E\sim1$ GeV, and higher energies,
$E\sim 100$ GeV.}
\label{tab:estbounds}
\end{table}
As a comparison, the ``most conservative'' bound for $n=1$ from Super-Kamiokande data
\cite{Fogli:1999fs} is $6 \times 10^{-24}$ at the 90\% confidence level.
The authors of \cite{Fogli:1999fs} comment that the strongest bounds are coming from
upgoing muon events, which peak at about 100 GeV, so their results are in line with
naive expectations.
The bounds of \cite{Fogli:1999fs} are improved by a factor of about 8 in the
combined analysis of Super-Kamiokande and K2K data \cite{Gonzalez-Garcia:2004wg}.
We would hope that the much greater number of such events that will be observed at a
neutrino telescope would enable these bounds to be improved.

\begin{figure}[h]
\begin{center}
\includegraphics[width=6.8cm]{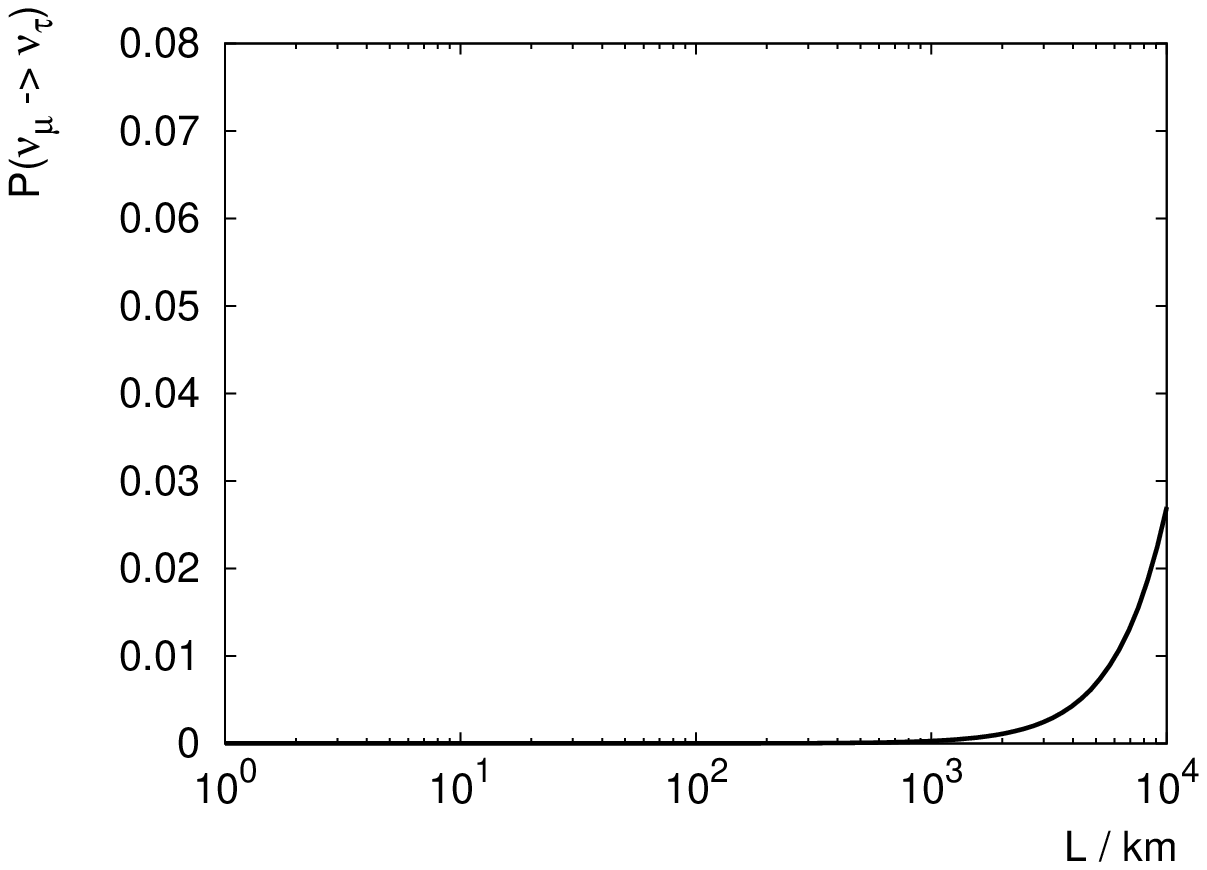}
\includegraphics[width=6.8cm]{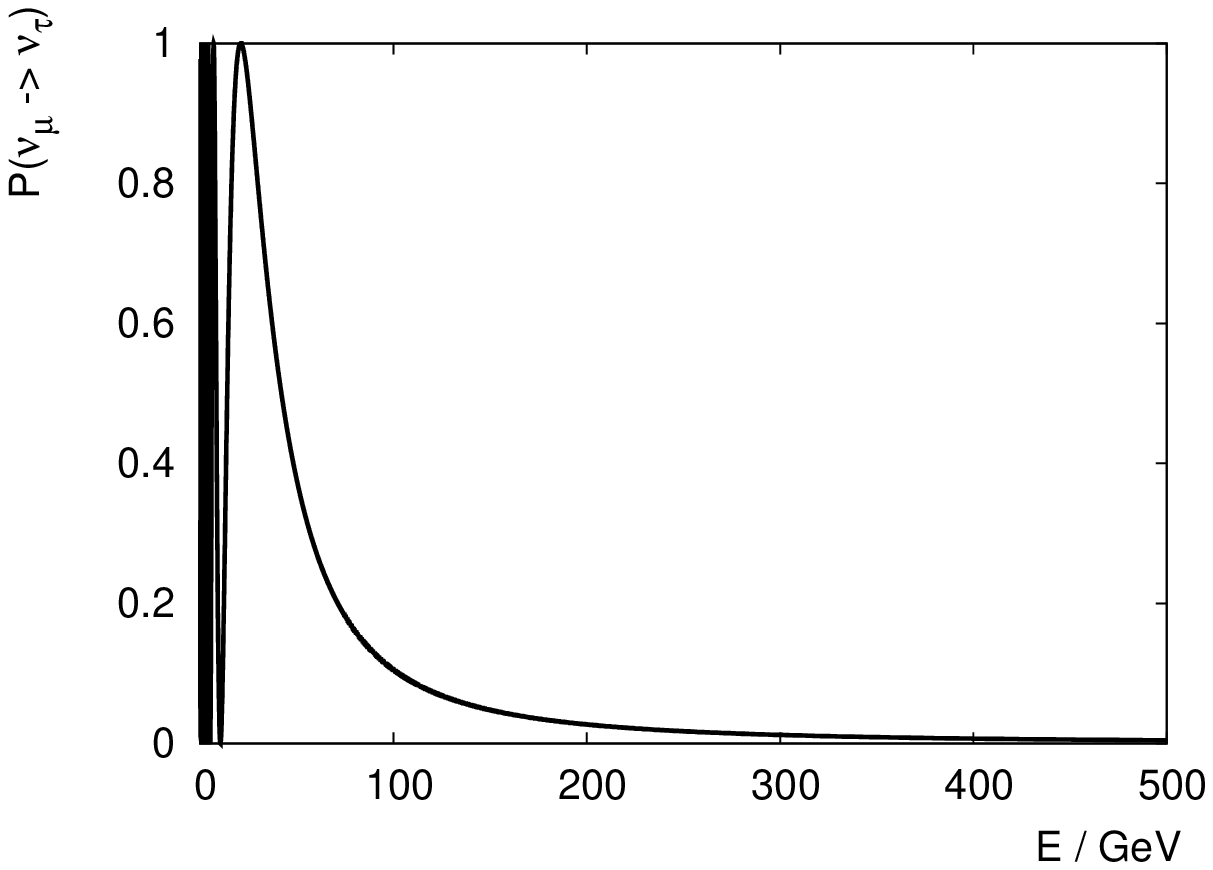}
\caption{Standard atmospheric neutrino oscillation probability
(\ref{lvtheory:eqn:standprob}).
The left plot shows the oscillation probability as a function of path length $L$
(measured in km), for fixed neutrino energy $E$ equal to 200 GeV.
The right plot show the oscillation probability as a function of energy for
path length $L=10^{4}$ km.}
\label{fig:standard1}
\end{center}
\end{figure}

To see how Lorentz invariance violation affects atmospheric
neutrino oscillations, it is useful to compare the
oscillation probabilities. Whilst the neutrino oscillation
probabilities themselves are not observables, they are useful
for gaining insight into the underlying physics.
Firstly, for comparison, we consider the standard
oscillation probability which is given by (\ref{lvtheory:eqn:standprob}).
In all the probabilities
presented here, we take \cite{Gonzalez-Garcia:2003qf}
\begin{displaymath}
\Delta m^{2}=2.6\times10^{-3}~{\rm{eV}}^{2}; \;\;\;\;\; \sin^{2}2\theta =
1.
\end{displaymath}
Fig. \ref{fig:standard1} shows the oscillation probability plotted
as a function of the path length, $L$, for neutrino energy
200 GeV and as a function of the neutrino energy, $E$,
for a fixed path length, $L=10^{4}$ km (which is comparable with the
diameter of the Earth, and used in the oscillation probability plots in
\cite{Giorgini:2005zd}).

\begin{figure}
\begin{center}
\includegraphics[width=6.8cm]{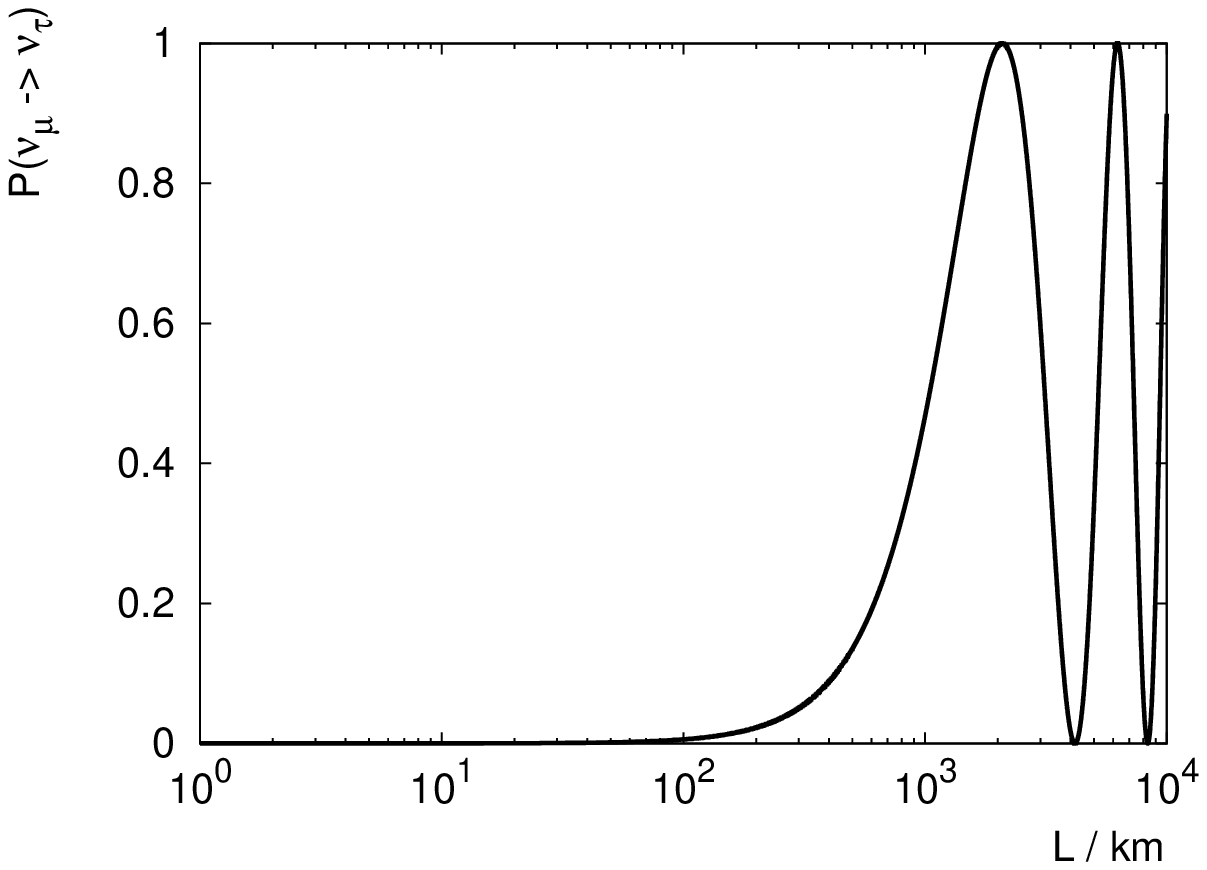}
\includegraphics[width=6.8cm]{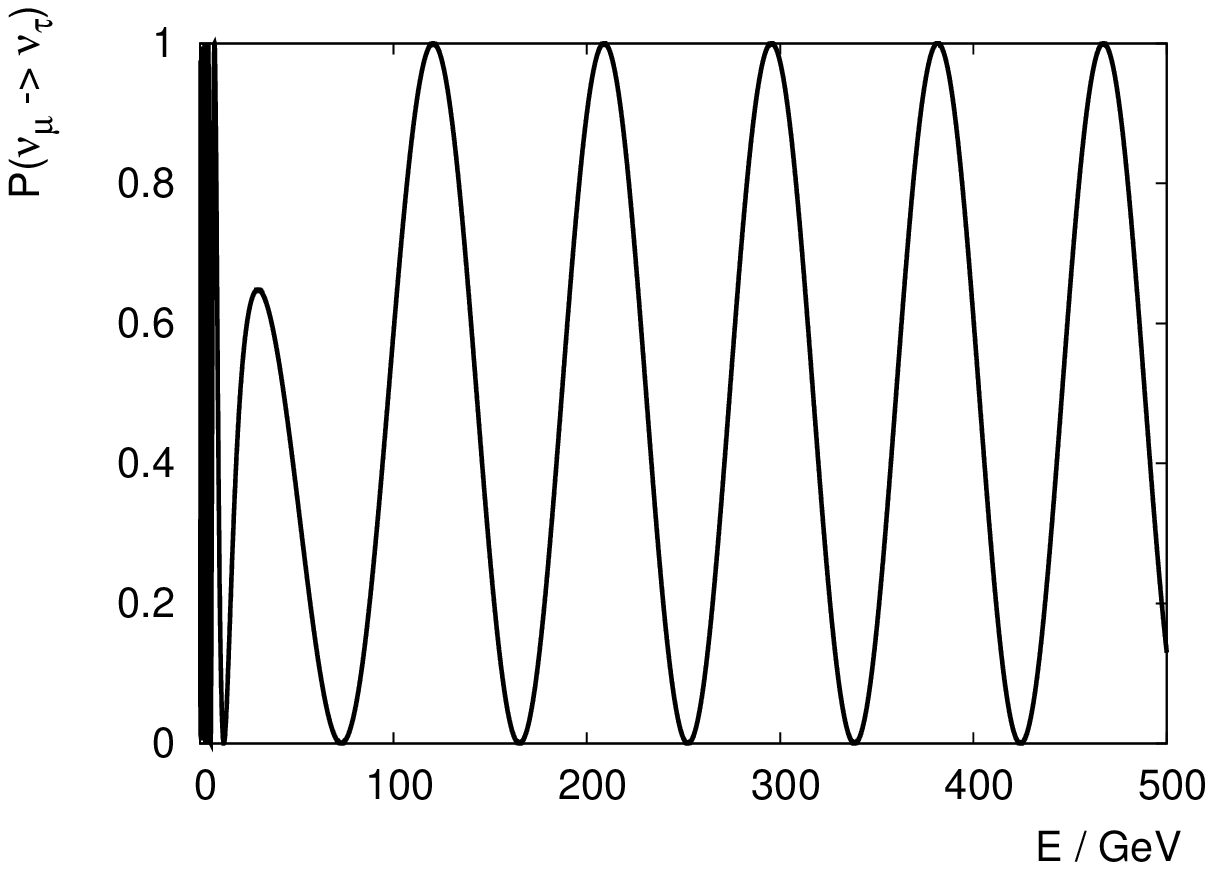}
\end{center}
\caption{Atmospheric neutrino oscillation probabilities for
$n=1$ ($\alpha=0$). The left plot shows the oscillation probability
as a function of path length for neutrinos with an
energy of $200$ GeV whilst the right plot shows the oscillation
probability as a function of energy for path length
$L=10^{4}$ km.}
\label{fig:2nprobn1}
\end{figure}

\begin{figure}
\begin{center}
\includegraphics[width=6.8cm]{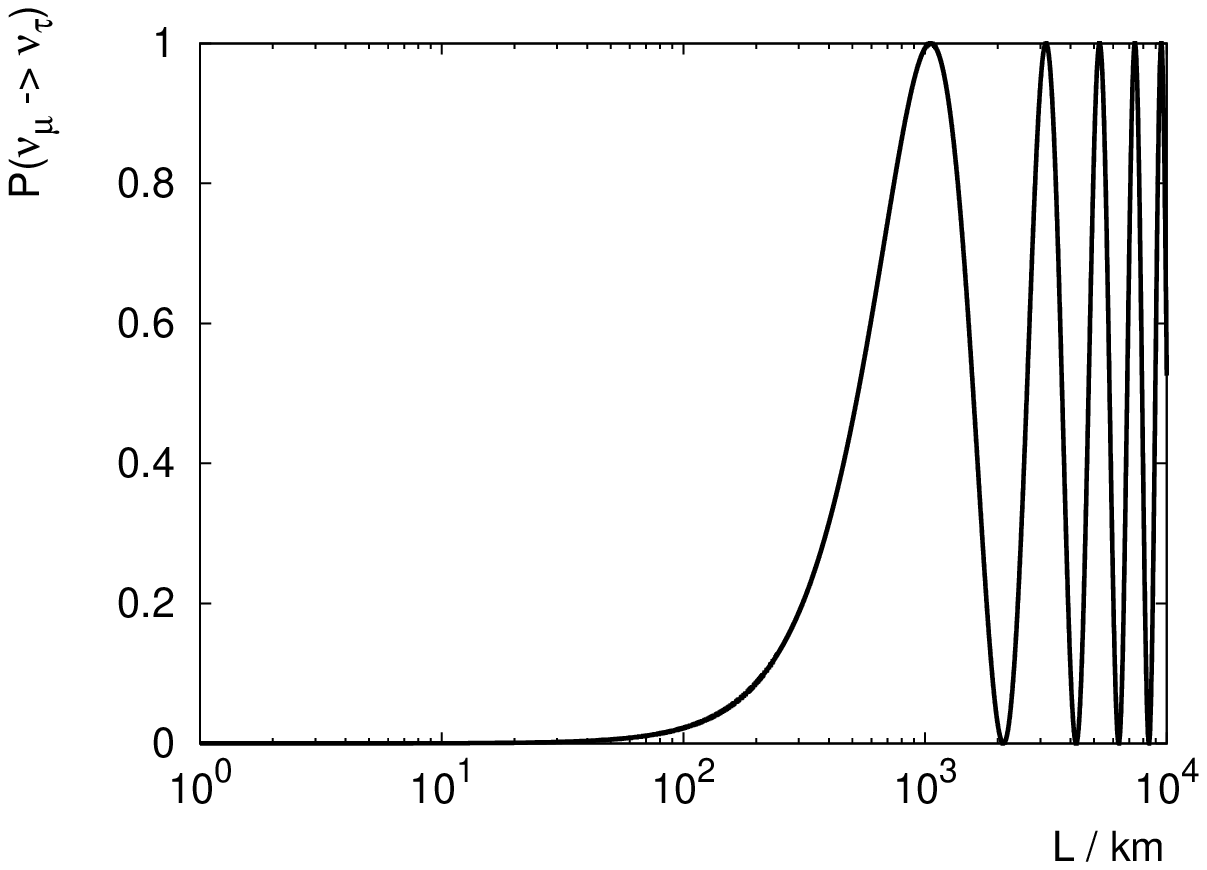}
\includegraphics[width=6.8cm]{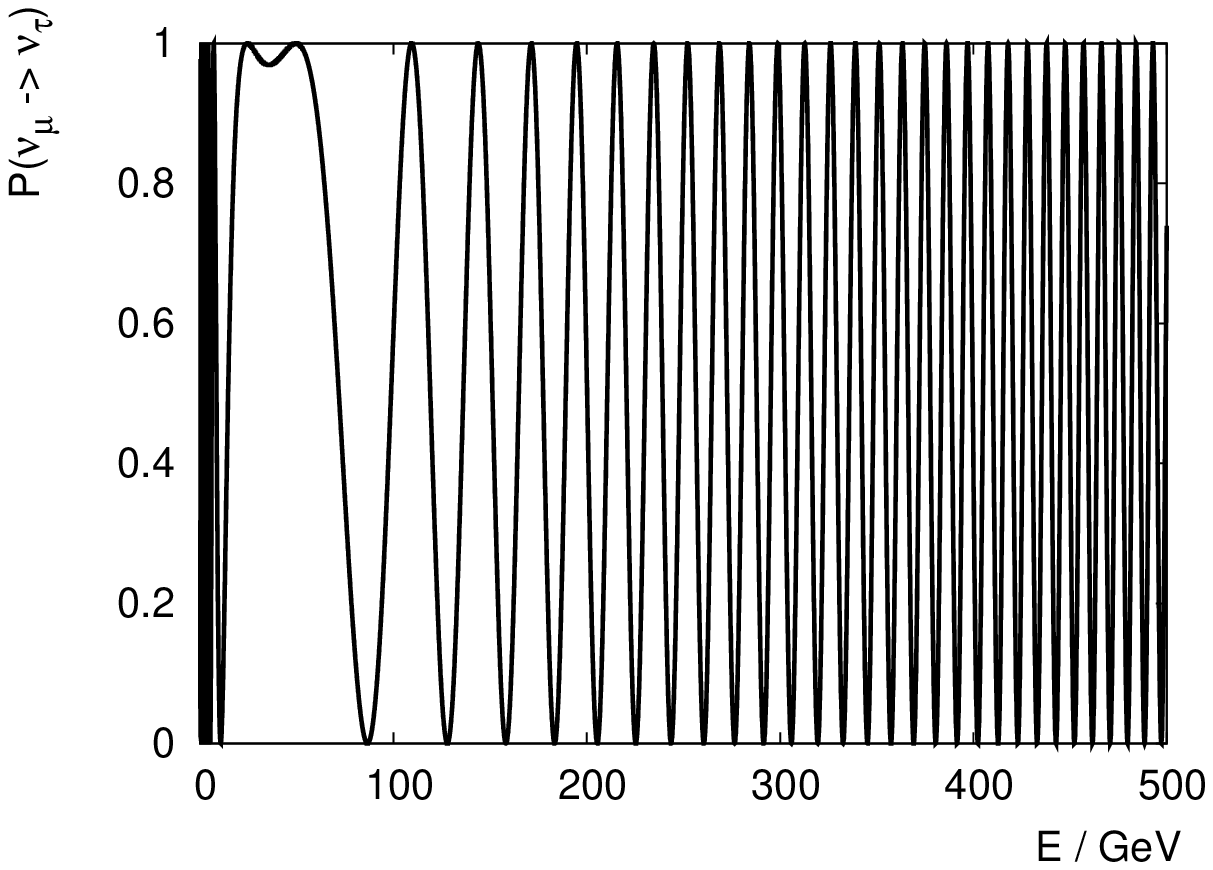}
\end{center}
\caption{As Fig. \ref{fig:2nprobn1} but
with $n=2$ ($\alpha=1$).}
\label{lvtheory:fig:2nprobn2}
\end{figure}

\begin{figure}
\begin{center}
\includegraphics[width=6.8cm]{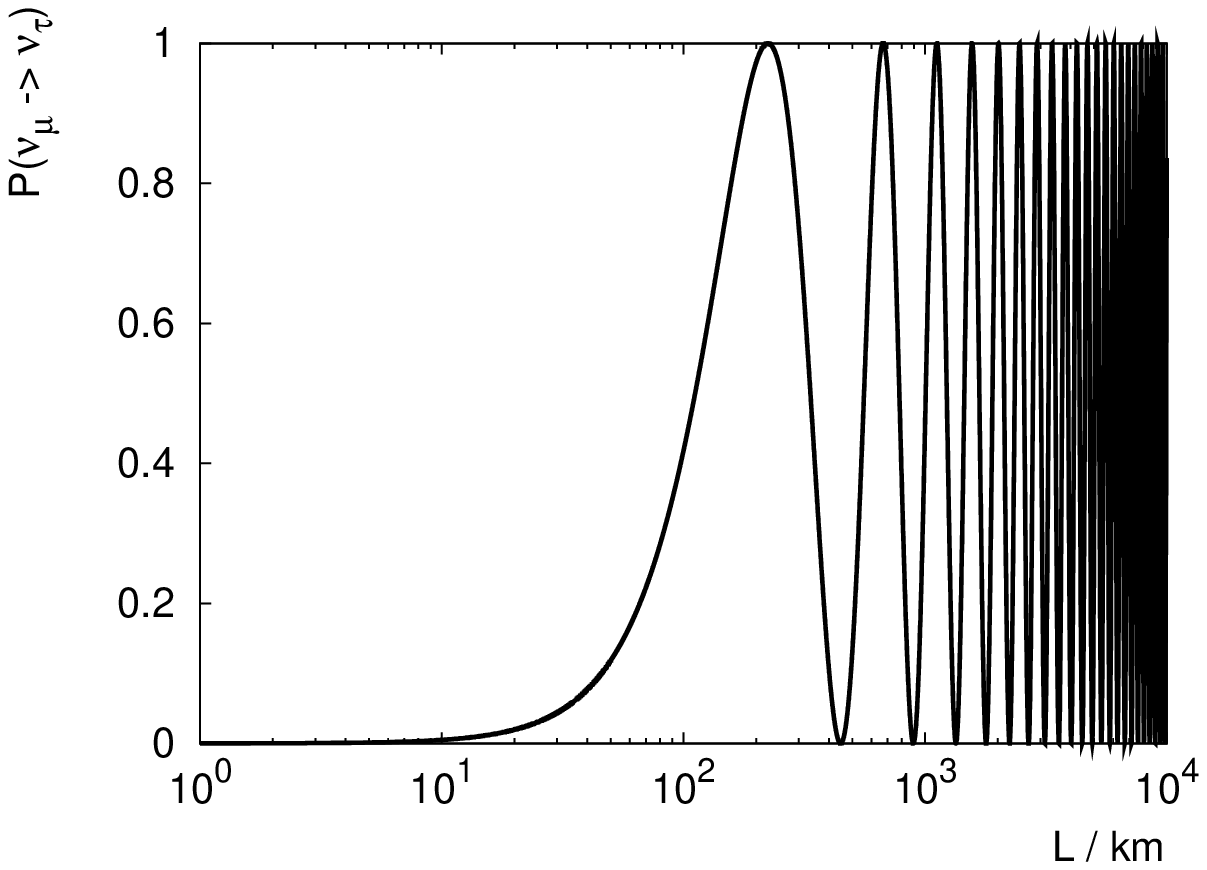}
\includegraphics[width=6.8cm]{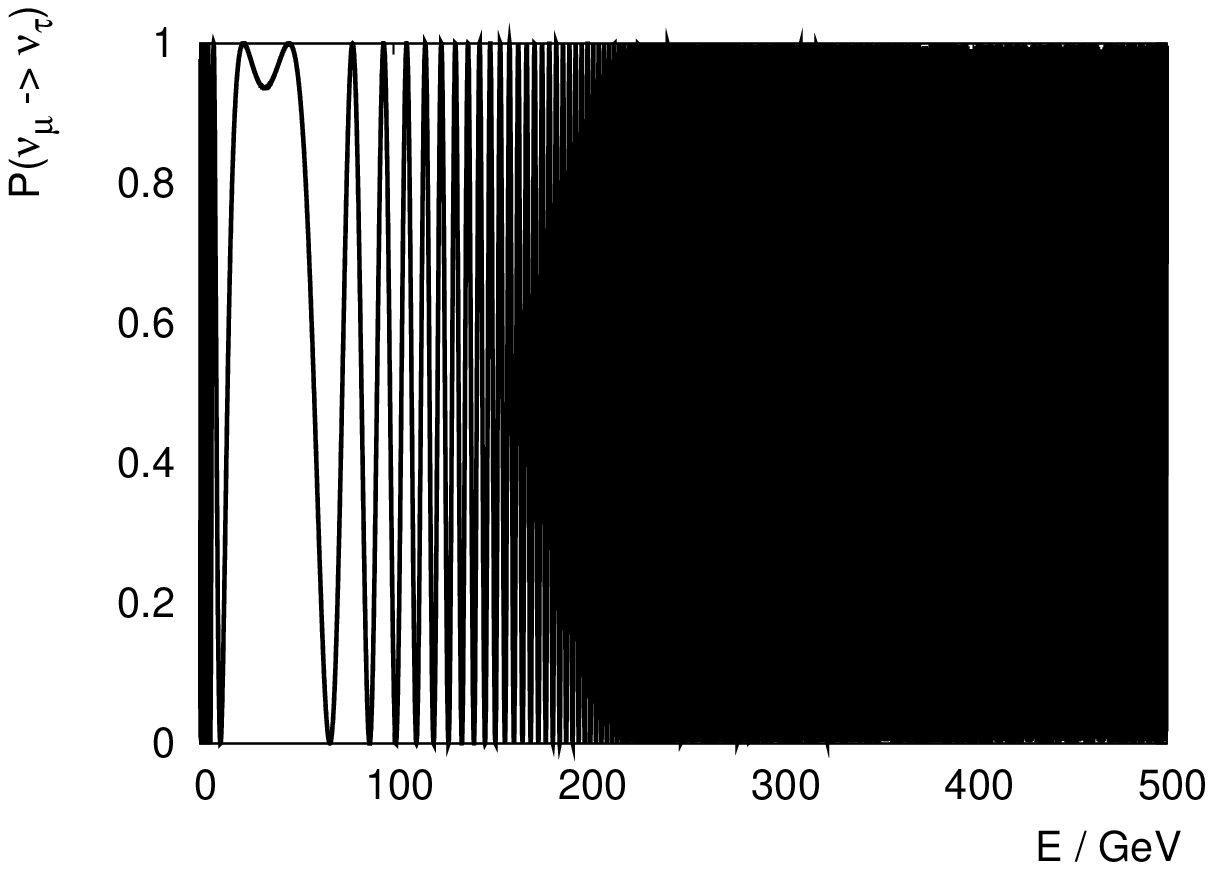}
\end{center}
\caption{As Fig. \ref{fig:2nprobn1} but
with $n=3$ ($\alpha=2$).}
\label{lvtheory:fig:2nprobn3}
\end{figure}

Figs. \ref{fig:2nprobn1}, \ref{lvtheory:fig:2nprobn2} and
\ref{lvtheory:fig:2nprobn3} show the oscillation probability
including LV effects (\ref{lvtheory:eqn:2nprobch}) for
$n=1,~2$ and 3 respectively. To construct these plots, we
used the upper bounds found in section \ref{sec:sim} and shown in table
\ref{lvsim:tab:diagbounds}.
In each case, we found that for low energy
neutrinos, the effects of LV on the neutrino oscillation probability are negligible.
However, for higher energy
neutrinos, this is not the case.
If we compare the plots of
probability against path length for high energy neutrinos, we see
that for standard oscillations, almost no oscillations occur,
however, if we include the LV effects, the oscillation probability
becomes significant at much shorter path lengths. Indeed, as $n$ increases,
the probability becomes significant at smaller path lengths. When we consider
the plots in which the probability is a function of energy, we see that for standard
oscillations, the probability tends towards zero, where as, in the plots which include
LV effects, the probability continues to oscillate with the period of oscillations decreasing
for increasing values of $n$.
Corresponding plots of atmospheric neutrino survival probabilities can be found in
\cite{Giorgini:2005zd} for the mass and velocity eigenstate mixing LV model.
Their plots have some features in common with ours, in particular the existence of
oscillations in the probability at high energies, when the survival probability
in the standard oscillation scenario has converged to unity.

In \cite{morgan:2004vv}, we studied in depth the effect of quantum decoherence on
atmospheric neutrino oscillations.
In that case, there were also marked differences in the oscillation probability, particularly
for high energies and long path lengths.
However, the probabilities with LV are rather different from those when we have quantum
decoherence.
In the decoherence case, the oscillations in the probability are damped (the rate of
damping depending on a number of factors), but here we find that the oscillations
in the probability continue for ever.
While these two possible sources of new physics give very different properties of the
oscillation probability, the key test is whether these differences can be observed
experimentally.
We turn to this issue in the next section.

\section{ANTARES sensitivity to LV in atmospheric neutrino oscillations}
\label{sec:sim}

The ANTARES neutrino telescope \cite{Korolkova:2004pg} is sensitive to high energy
neutrinos above the energy threshold of 10 GeV. These
neutrinos do not form the primary source of neutrinos for
ANTARES, instead they form the background to
astrophysical and cosmological neutrinos.

\subsection{Simulations}

Our simulation strategy follows that presented in \cite{morgan:2004vv},  so
we only briefly outline the key features.
Further details are documented in \cite{morgan:2004vv}, and in
the analysis of standard atmospheric neutrino oscillations
\cite{Brunner:1999ig}, on which our
simulations are based (with suitable modifications for the inclusion of LV).

Atmospheric neutrino events were generated using a Monte-Carlo (MC)
production assuming a spectrum of $E^{-2}$ for neutrinos in
the energy range $10~{\rm{GeV}}<E<100~{\rm{TeV}}$. The angular
distribution was assumed to be isotropic. A total of twenty
five years of data were simulated in order that errors from
the Monte-Carlo statistics could be ignored and event
weights are used to adapt the Monte-Carlo flux to a real
atmospheric neutrino flux. The Bartol theoretical flux
\cite{Gaisser:2001jw} was
used here although we do not anticipate the results
presented will change using a different flux, however,
there may be a significant change to our results if a
different spectral index is assumed.

The simulations presented here are based on three years' worth of
data taking with ANTARES, corresponding to around 10000
atmospheric neutrino events. All errors are purely
statistical with simple Gaussian errors assumed. We are able
to produce spectra in either $E$, $E/L$ or $L$ and, by
altering the oscillation probability in the neutrino
oscillation code, we are able to include LV effects.
The path length $L$ is related to the zenith angle $\vartheta $
(not to be confused with the mixing angle $\theta $)
as follows, provided $\cos \vartheta \neq 0$:
\begin{displaymath}
L \approx 2R \cos \vartheta ,
\end{displaymath}
where $R = 6378$ km is the radius of the Earth,
and $\vartheta = 0$ corresponds to a neutrino travelling vertically upwards.
We are therefore interested in typical path lengths of the order $L\approx 10^{4}$ km.
ANTARES is particularly sensitive to the zenith angle
$\vartheta $ and hence the path length $L$.
In the sensitivity analysis, we therefore used the spectra in $E/L$.
However, the spectra in $L$ alone also have some interesting features,
which we discuss below.

In order to produce sensitivity regions, we use a $\chi^{2}$
technique which compares the $\chi^{2}$ for oscillations with
LV, or LV effects alone, with that for the no-oscillation
hypothesis. We leave the total normalization as a free
parameter so that our results are not affected by  the
normalization of the atmospheric neutrino flux. Sensitivity
regions at both the 90 and 99 percent confidence level are
produced.

Although our analysis is independent of the normalization of
the atmospheric neutrino flux, in order to produce spectra
based on the number of observed events, some normalization
is necessary. For standard oscillations with no LV effects,
bins at high $E/L$ can be used to normalize the flux as
standard oscillations become negligible here. However, since
all our LV models are proportional to positive powers of the
neutrino energy, LV effects are significant at high energies
but negligible at low energies, and thus low $E/L$. In this
case, we use the low $E/L$ bins to normalize the flux.

In our simulations, we consider cases where $\Delta m^{2}$
is zero, so neutrino oscillations arise due to LV
effects only, and where $\Delta m^{2}$ is non-zero, so that the LV effects modify
standard neutrino oscillations. In the case when $\Delta
m^{2}$ is non-zero, we also allow this parameter to vary in
order to check whether the experimental point of best fit
lies inside our sensitivity regions. In
this case, we have three parameters, $\Delta m^{2}$, $\Delta
\eta$ and $\sin^{2}2\theta$. We therefore obtain a three
dimensional sensitivity region at 90 and 99\% confidence
levels. In order to consider the details of the sensitivity
region, we plot the boundaries of the projection of the
surface bounding this volume onto the coordinate planes,
thus obtaining three distinct plots.

\subsection{Spectra}

In addition to data based on the total number of observed
events, the spectra of these events, typically in $E/L$, is
also important. For example, spectral analysis
has been performed for Super-Kamiokande \cite{Fogli:1999fs,Gonzalez-Garcia:2004wg,Ashie:2004mr},
KamLAND \cite{Araki:2004mb} and K2K \cite{Gonzalez-Garcia:2004wg,Aliu:2004sq}. In
addition to being used in the construction of sensitivity
regions, the spectra themselves are also important for
bounding the size of possible LV effects and possibly ruling
effects out due to the observed shape of the spectra.
As explained above, we used the spectra in $E/L$ in the sensitivity
analysis, so we focus on those spectra (apart from Fig.
\ref{lvsim:fig:LVspectraLVn=1}).

\subsubsection{Model LV1}
\label{subsec:lv1}

We begin by examining the spectra obtained when we consider the
possibility that neutrino oscillations occur as a consequence of
LV only ($\Delta m^{2}=0$ in Eq. (\ref{lvtheory:eqn:2nprobch})),
when the LV effects are proportional
to the neutrino energy, $n=1$.
\begin{figure}
\begin{center}
\includegraphics[width=6.8cm]{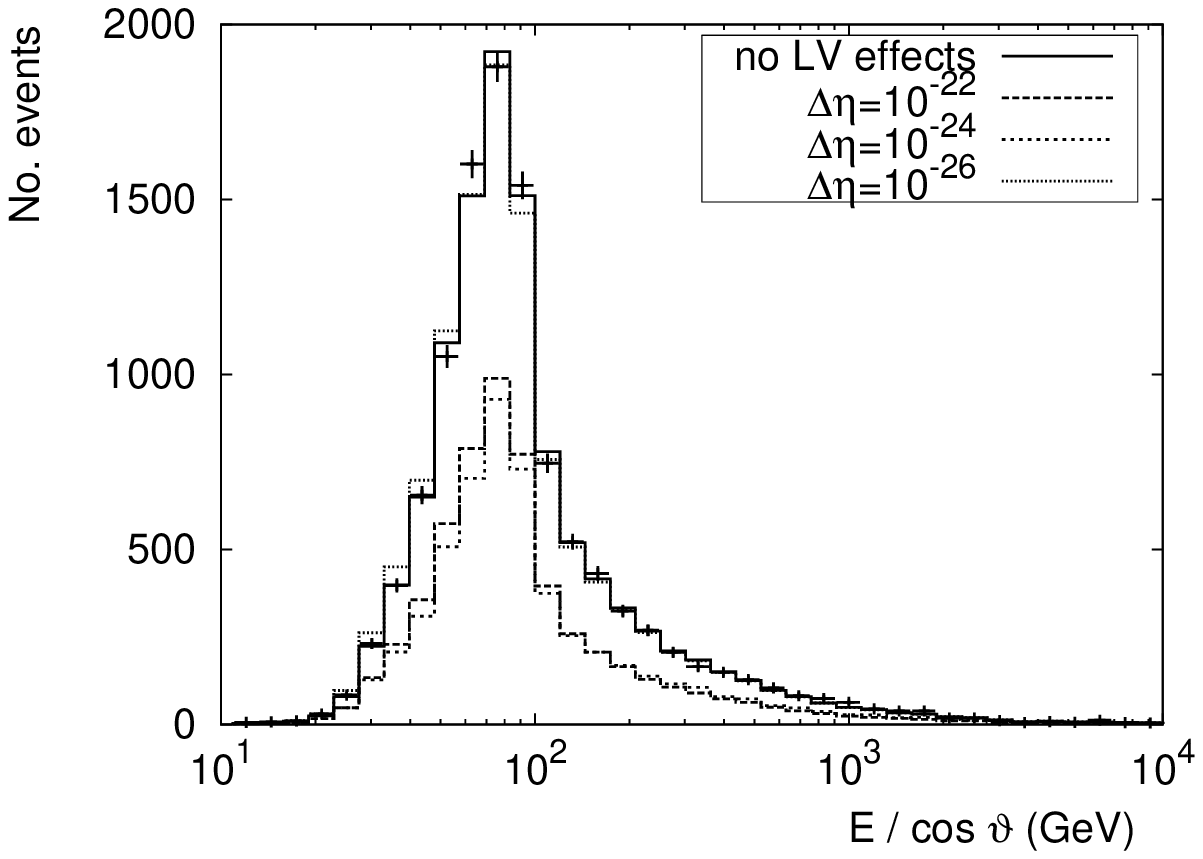}
\includegraphics[width=6.8cm]{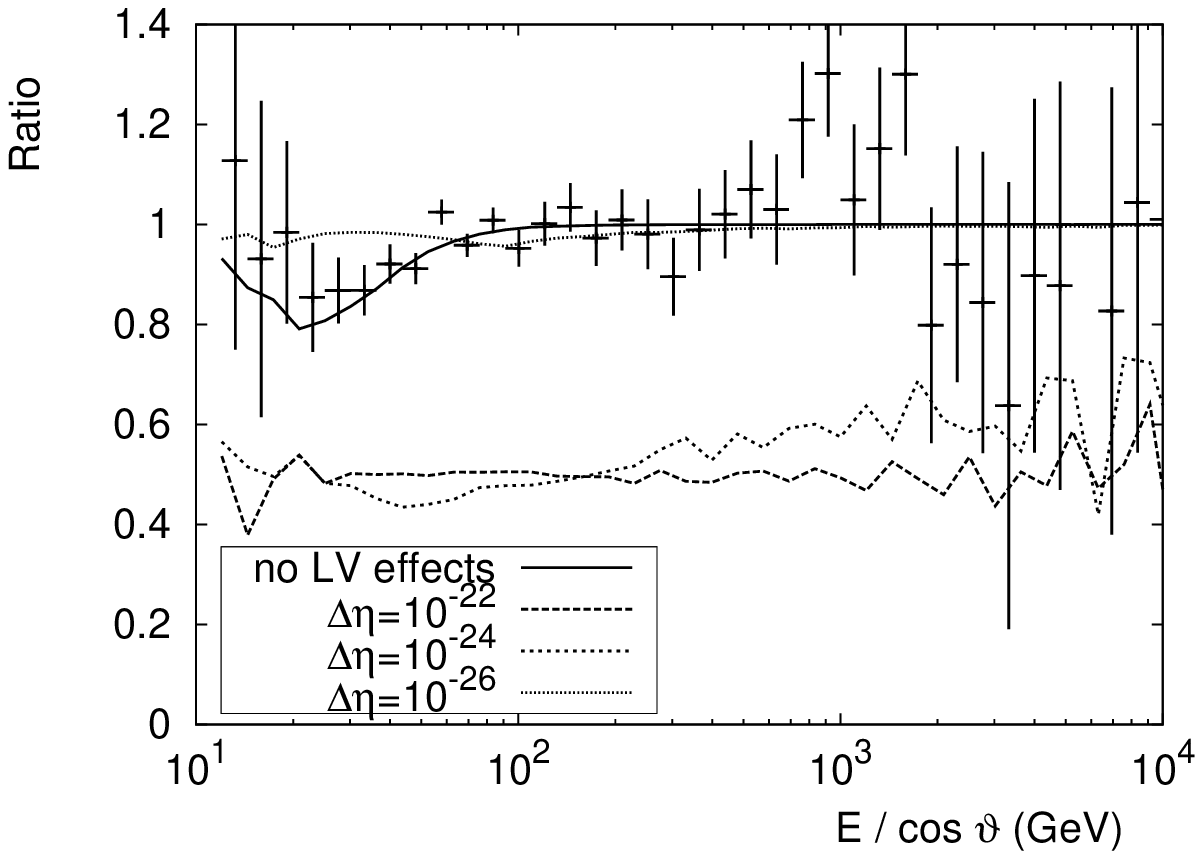}\vspace{2cm}
\includegraphics[width=6.8cm]{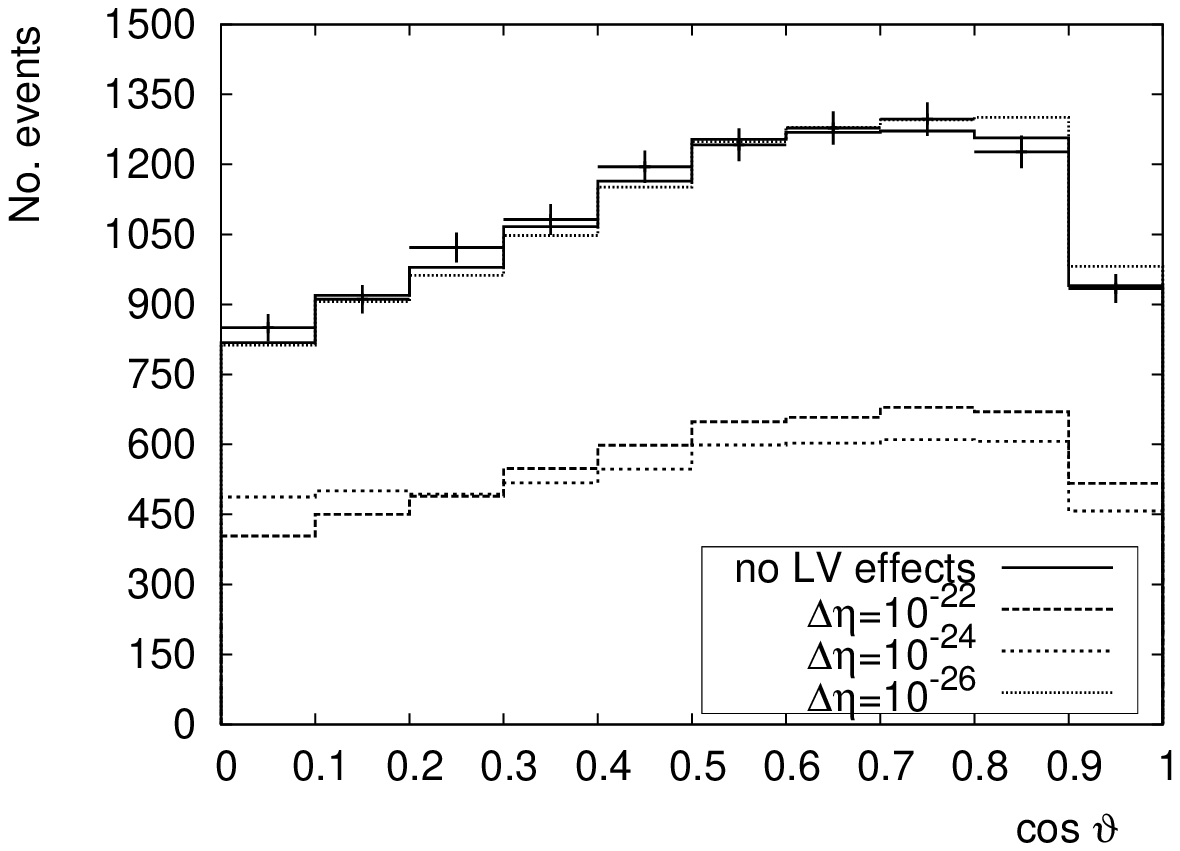}
\includegraphics[width=6.8cm]{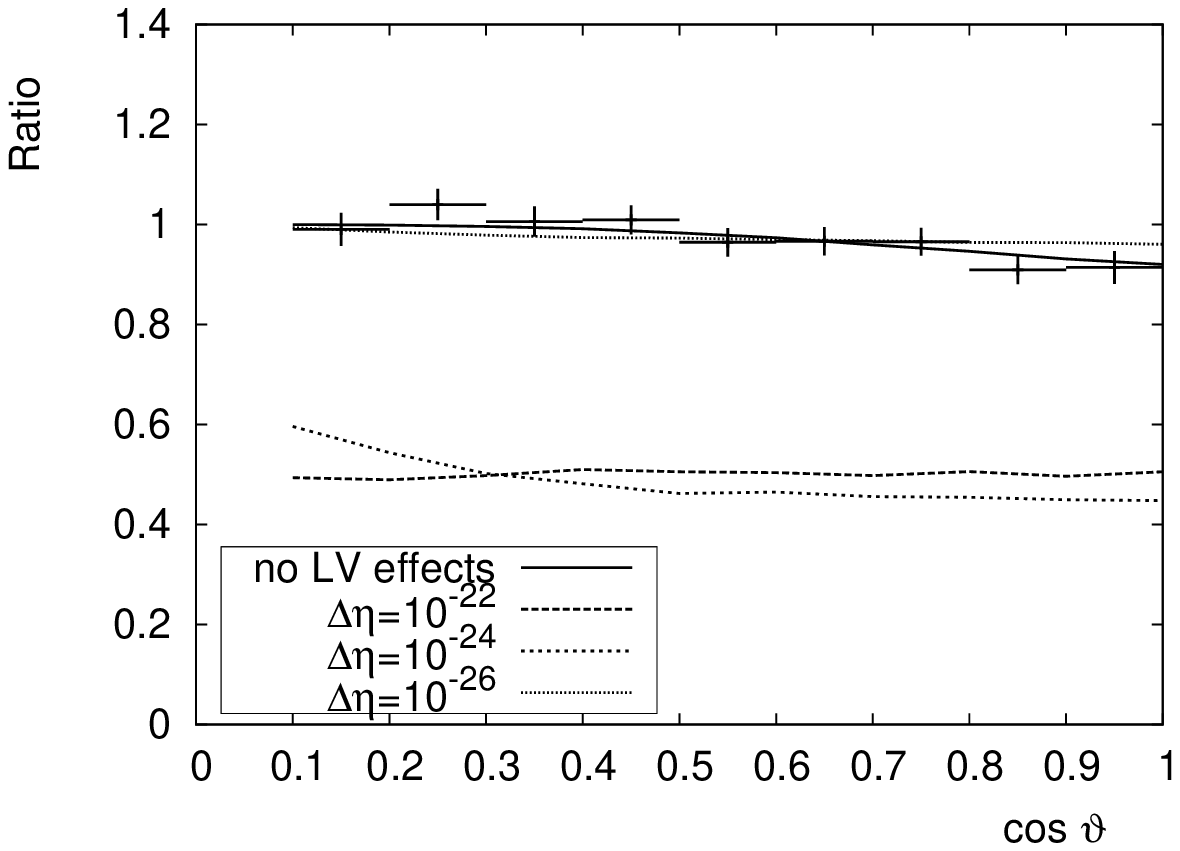}
\end{center}
\caption{Spectra of events for atmospheric neutrino oscillations due to LV effects only,
when the  LV
effects are proportional to the neutrino energy (left panel)
and their ratios to the events without oscillations (right panel)
as functions of $E/\cos \vartheta $ (upper plots) and $\cos \vartheta $ (bottom plots).
The solid line
represents the MC simulation of standard oscillations with the
dotted/dashed lines showing the spectra for oscillations from
LV only.
The simulated data points correspond to three-years' data taking, assuming
standard oscillations only with no LV.}
\label{lvsim:fig:LVspectraLVn=1}
\end{figure}
Fig. \ref{lvsim:fig:LVspectraLVn=1} shows
the expected spectra of events as a function of $E/\cos\vartheta$
and $\cos\vartheta$.
The solid line corresponds to the MC simulation with
standard oscillations whereas the dotted/dashed lines are the
MC simulations for LV only induced oscillations.
In each case, we have also plotted simulated data points, based on the
standard oscillation picture with no LV effects.
On the right-hand-side, we show
the ratios of the number of
events compared to no oscillations.

The spectra show that,
for the larger values of
the LV parameter, there is a definite suppression in the number of
observed events by a factor of approximately two. If, however, we
consider the case in which the LV parameter takes the smaller value,
it is more difficult to distinguish oscillations that arise as a
consequence of LV effects from those arising in the standard
oscillation picture. If we consider the
ratios,
we note that there are some features which would
allow us to differentiate between the two phenomenologies. In the
top right frame, we note that at low energies, the LV model does
not result in an oscillation minimum whereas the standard
oscillation model does. Secondly, although the difference is
small, the bottom right frame shows a deviation from the standard
picture at high values of the zenith angle corresponding to large
path lengths.
We will not consider further spectra just in the path length $\cos \vartheta $,
but note that for all the models we
study and
all the energy dependences of the LV parameters, there are slight deviations in the
shape of the spectra in $\cos \vartheta $ compared with the standard oscillation case.
In the analysis of Super-Kamiokande data \cite{Fogli:1999fs,Gonzalez-Garcia:2004wg}
it is the spectra in the path-length $L$ which are used to bound the magnitudes of
LV parameters, as the inclusion of LV effects again alters the shape of the spectra.

For the remainder of this section, we focus on the ratios of the numbers of events
in each case compared with the expected number of events if there were no oscillations.
\begin{figure}
\begin{center}
\includegraphics[width=10cm]{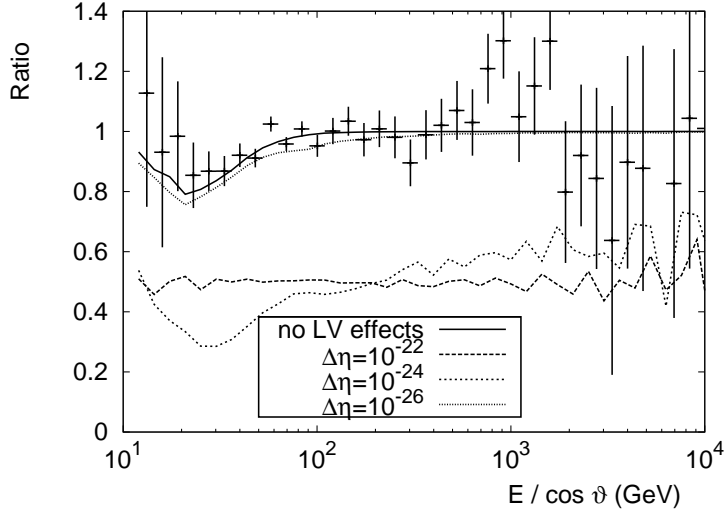}
\end{center}
\caption{Ratio of the number of events for standard oscillations modified
by LV effects proportional to the neutrino energy, compared with no oscillations.
The solid line
represents the MC simulation of standard oscillations with the
dotted/dashed lines showing the spectra for standard oscillations modified by
LV.
The simulated data points assume only standard oscillations, with no LV effects.}
\label{lvsim:fig:LVspectran=1}
\end{figure}
Fig. \ref{lvsim:fig:LVspectran=1} shows the
ratios in the case of standard oscillations plus LV
as a function of $E/\cos\vartheta$. Again, we
note that large values of the LV parameters result in a large
suppression of the number of expected events. Also, for very large
values of the LV parameter, we see a considerable flattening in
the ratio of the number of events compared with no oscillations.
The situation is much trickier, however, for very small values of
the LV parameter. In this case, the number of events expected is
not suppressed  and we also see an oscillation minimum. Therefore,
this makes it very difficult to distinguish between standard
oscillations and those which are modified to include LV effects,
if the LV parameter is of the order of $10^{-26}$ or smaller.

\subsubsection{Models LV2 and LV3}
\label{subsec:lv2}

\begin{figure}
\begin{center}
\includegraphics[width=10cm]{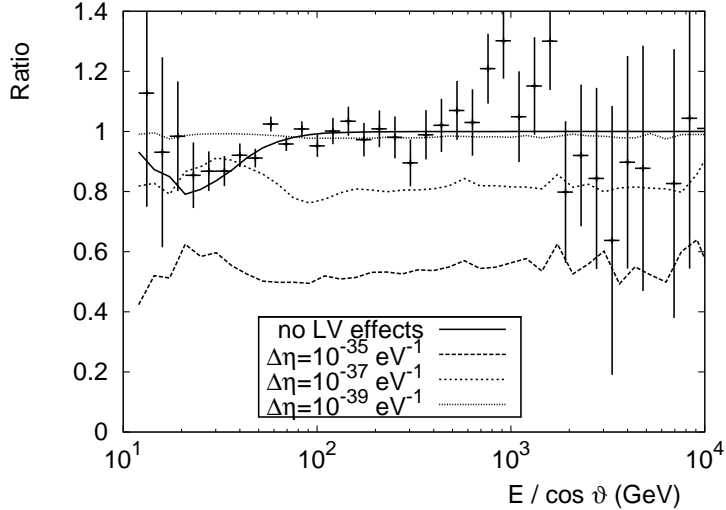}
\end{center}
\caption{Ratio of the number of events for oscillations due to LV effects only,
 when the
LV effects are proportional to the neutrino energy squared, compared with no
oscillations. The
solid line represents the MC simulation of standard oscillations
with the dotted/dashed lines showing the spectra for oscillations from
LV only.}
\label{lvsim:fig:LVspectraLVn=2}
\end{figure}

We first
consider the spectra when the LV parameters are
proportional to the square of the neutrino energy. Fig.
\ref{lvsim:fig:LVspectraLVn=2} shows the ratio of
number of events with the LV effects only to those expected if there were no oscillations.
For large values of the LV parameters, we note that the spectra
are very different from standard neutrino oscillations. Firstly,
as in the case before, the number of events is significantly
suppressed for very large values of the LV parameter.
For
standard neutrino oscillations, we observe an oscillation minimum,
however, for large values of the LV parameter, the ratio increases
and so we see a peak. Even for smaller values of the LV parameter,
the ratio is flat and so by considering the spectra it should be
possible to distinguish between these two models.

\begin{figure}
\begin{center}
\includegraphics[width=10cm]{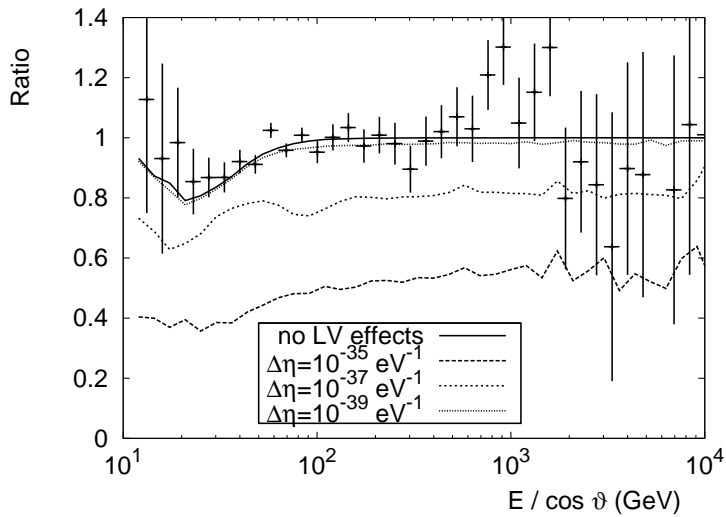}
\end{center}
\caption{As Fig.
\ref{lvsim:fig:LVspectraLVn=2} but the dotted/dashed lines show the
spectra for standard oscillations plus
LV.}
\label{lvsim:fig:LVspectran=2}
\end{figure}

Fig. \ref{lvsim:fig:LVspectran=2} shows the spectra in
the case where the LV effects modify standard
oscillations, then, as in  subsection \ref{subsec:lv1} before, it becomes much more
difficult to untangle any LV effects from those of standard
oscillations for small LV parameter.
For large values of the LV parameter, the
only distinguishing feature is that the number of events we expect
to see is somewhat reduced.
In addition, for large values of the LV parameter, the curve
is much flatter than we would expect from
standard oscillations.

We find qualitatively similar results when the LV effects are proportional
to the cube of the neutrino energy, both in the case where oscillations are due
to LV effects only, and when LV effects modify standard neutrino oscillations.
For both models LV2 and LV3, the shape of the spectra in the zenith angle
$\cos \vartheta $ only are slightly different to that shown in for model LV1 in Fig.
\ref{lvsim:fig:LVspectraLVn=1}.
The ratios of the number of events compared with no oscillations are flatter than in the
bottom right-hand plot in Fig. \ref{lvsim:fig:LVspectraLVn=1}, particularly for model LV3.
We do not use these spectra for the following sensitivity analysis, but it may be possible,
with real data, to use those spectra as well as those for numbers of events in terms of
$E/ \cos \vartheta $.

\subsection{Sensitivity regions}

In the previous section, we saw how LV effects can modify the
number of events seen in the detector and how they would modify
the spectra. We now turn to the discussion of the
ANTARES sensitivity regions found from our numerical simulations.
As in the case of quantum decoherence
\cite{morgan:2004vv}, we are particularly
interested in finding upper bounds for the LV parameters.

\subsubsection{Models LV1 and LV2}

In our first model, the LV effects are proportional to the neutrino energy.
We firstly consider the case in which $\Delta m^{2}=0$, so that
there are no standard oscillations, only those which arise as a
consequence of the LV effects.
\begin{figure}
\begin{center}
\includegraphics[width=7.5cm]{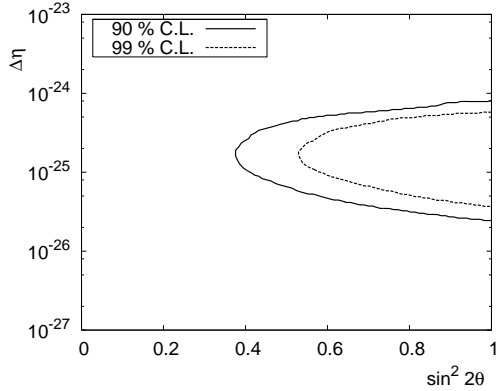}
\end{center}
\caption{Sensitivity contours at 90 and 99\%
confidence levels for LV effects only (no standard oscillations)
for model LV1, when the LV effects are proportional to the neutrino energy.}
\label{lvsim:fig:LVsenseLV1}
\end{figure}
Fig. \ref{lvsim:fig:LVsenseLV1}
shows the sensitivity curves in this case. It is clear
that the smallest values of the LV parameter may be probed when
$\sin^{2}2\theta$ is close to one, which is the region of parameter space
in which we are most interested. We are able to place an upper
bound on $\Delta \eta $ of $8.2\times10^{-25}$.

We also considered the case of non-zero $\Delta m^{2}$, when standard
atmospheric neutrino oscillations are modified by LV effects.
\begin{figure}
\begin{center}
\includegraphics[width=6.8cm]{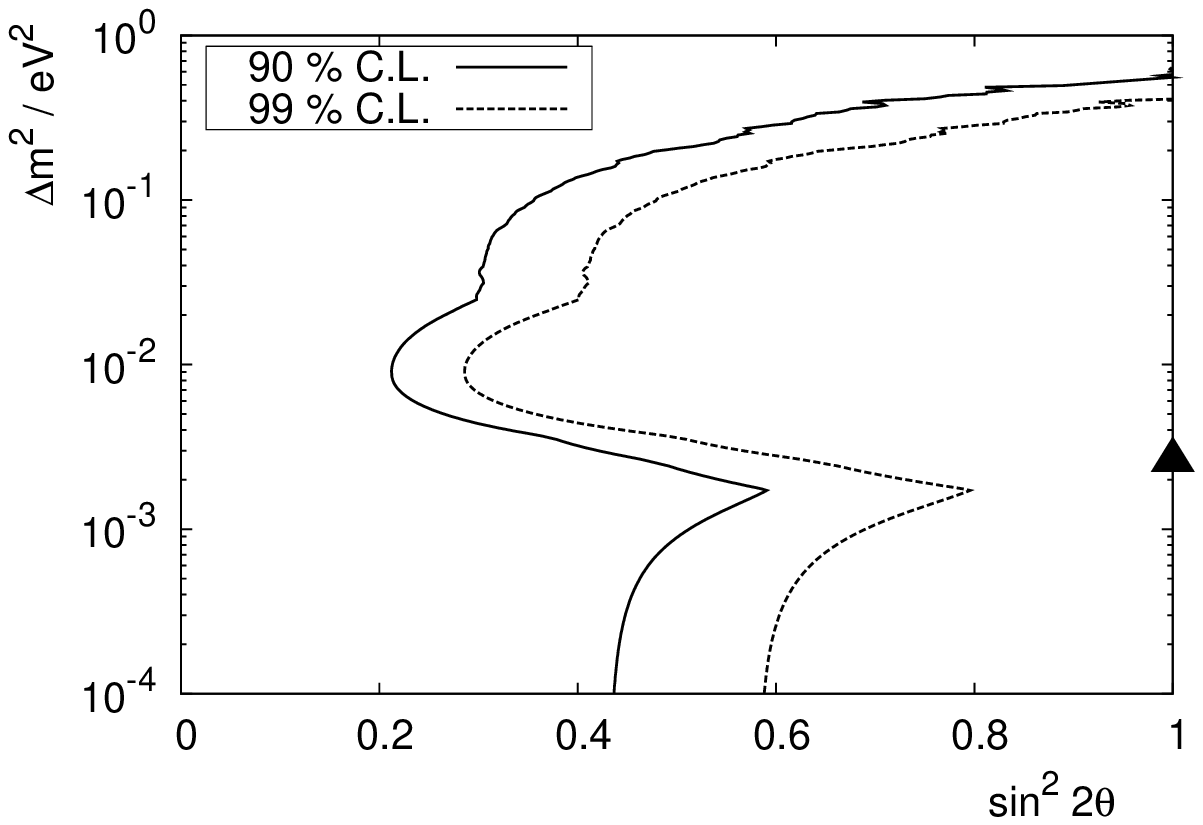}
\includegraphics[width=6.8cm]{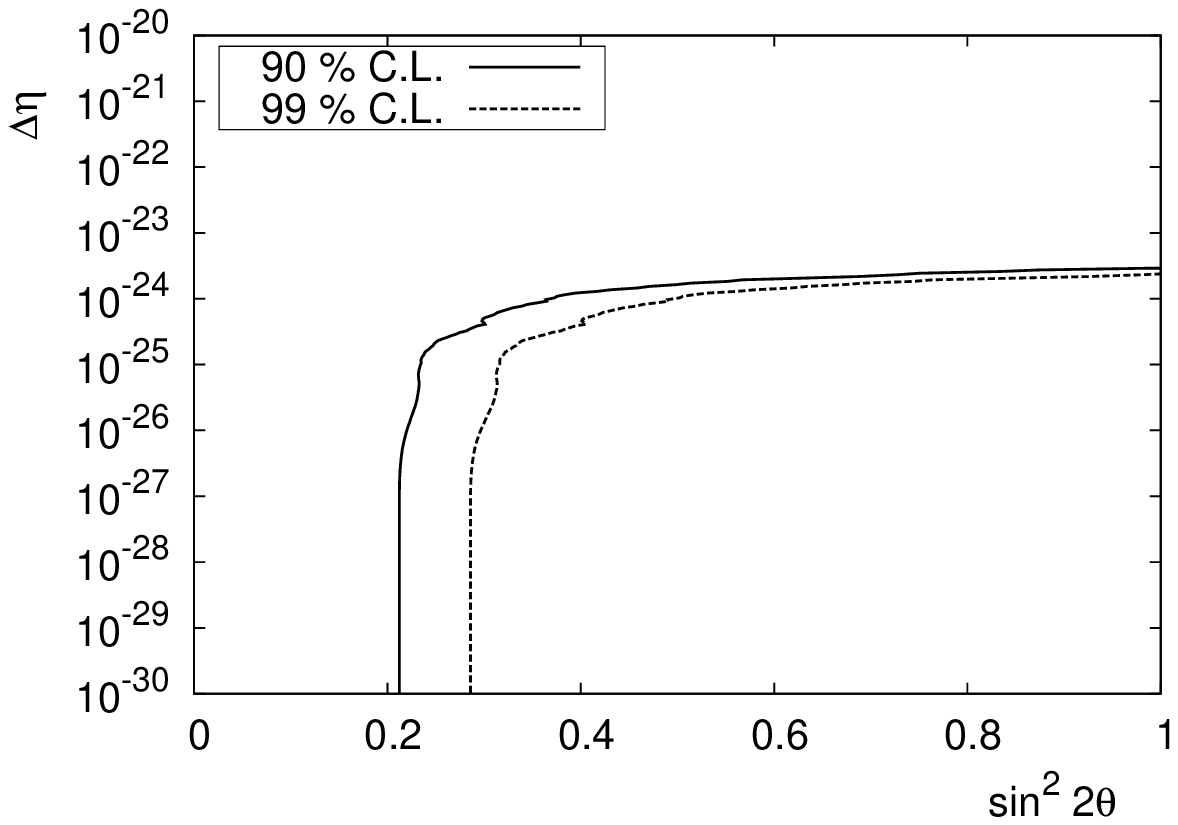}\vspace{2cm}
\includegraphics[width=6.8cm]{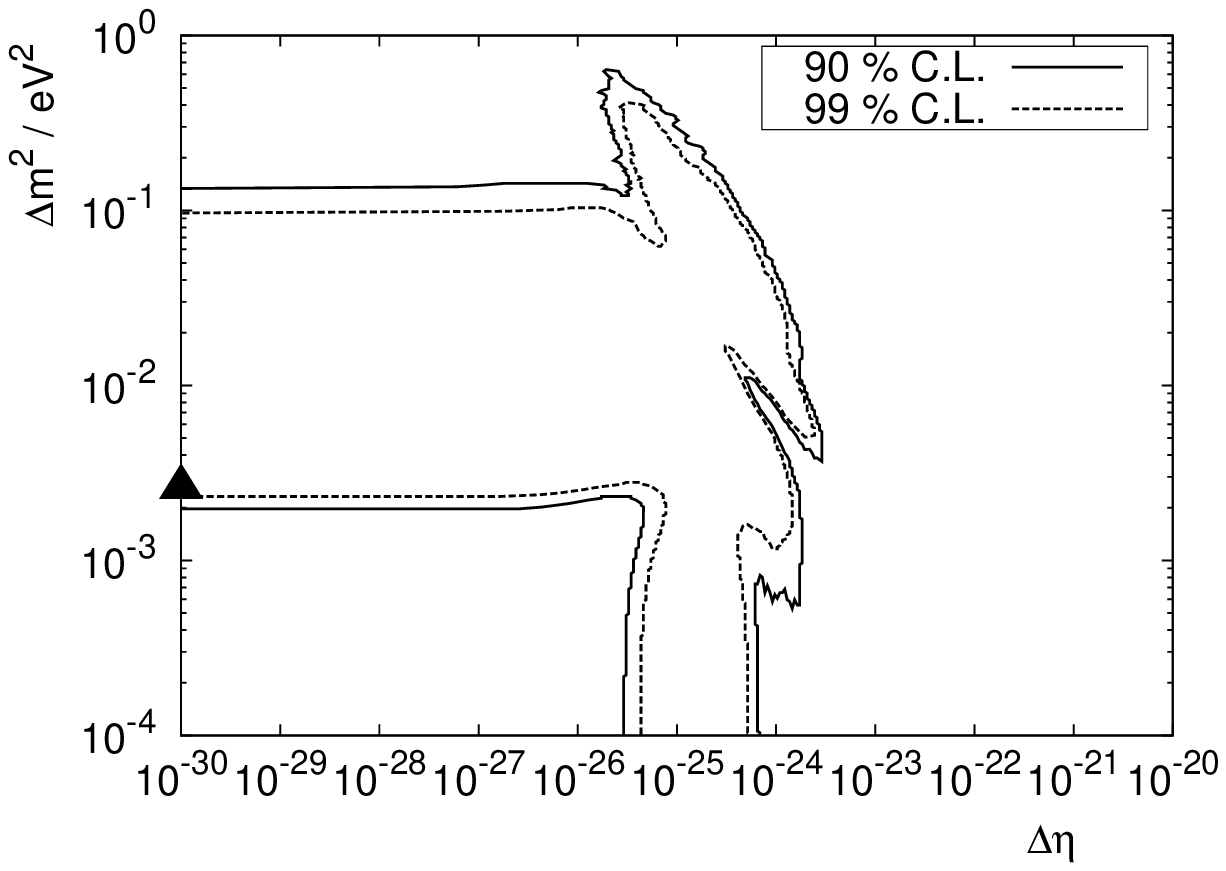}
\end{center}
\caption{Sensitivity contours for model LV1 at
90 and 99\% confidence level for standard oscillations plus
LV effects proportional to the neutrino energy.
The triangle denotes the best-fit value
of $\Delta m^{2}$ with no LV effects \cite{Gonzalez-Garcia:2003qf}.}
\label{lvsim:fig:LVsensen=1}
\end{figure}
Fig. \ref{lvsim:fig:LVsensen=1} shows the sensitivity
contours in this case.
Since we have varied three parameters ($\sin ^{2} (2\theta) $,
$\Delta m^{2}$ and $\Delta \eta $), we obtain a sensitivity volume.
Projecting the boundary surface of this volume onto the co-ordinate planes gives us
the sensitivity contours plotted in Fig. \ref{lvsim:fig:LVsensen=1}.
As we see from Fig. \ref{lvsim:fig:LVsensen=1},
the experimental point of best fit of $\Delta m^{2}$
\cite{Gonzalez-Garcia:2003qf}, denoted by the triangle, lies
within the sensitivity region. However, the region extends down to
very small values of $\Delta m^{2}$ and so we find that our
results include the previous case where neutrino oscillations arise
as a consequence of LV effects only. The top right frame in Fig.
\ref{lvsim:fig:LVsensen=1} shows the sensitivity contour for the
$\Delta \eta $ as a function of the mixing angle, $\theta$.
Again, $\Delta \eta =0$ is in this region, and so standard
oscillations with no LV effects are also included,
with an upper limit shown in Table \ref{lvsim:tab:diagbounds}.
\begin{table}[h]
\center{
\begin{tabular}{|c|c|c|c|}
\hline
$n$  & $\Delta\eta~({\rm{eV}}^{1-n})$  & $\Delta\eta~({\rm{eV}}^{1-n})$
& Planck scaling \\
& ($\Delta m^{2}=0$) & ($\Delta m^{2}\ne0$)&  (eV$^{1-n}$)\\
\hline
1 & $8.2\times10^{-25}$ & $2.9\times10^{-24}$ &
X \\
\hline 2 & $1.0\times10^{-35}$ & $2.9\times10^{-35}$ &
 $10^{-28}$ \\
\hline 3 & X & $6.9\times10^{-46}$  &
 $10^{-56}$ \\
\hline
\end{tabular}}
\caption{Table showing the upper bounds on $\Delta\eta$
contained with the sensitivity regions for various values of $n$.
We also give the naive expected values of $\Delta \eta $ from suppression by
appropriate powers of the Planck mass.
The X indicates that we were unable to place a bound on this
parameter.}
\label{lvsim:tab:diagbounds}
\end{table}
This is the same order of magnitude as the upper bound from Super-Kamiokande and K2K
data \cite{Fogli:1999fs,Gonzalez-Garcia:2004wg} so it seems unlikely that ANTARES
will be able to significantly improve on current data.

We have performed a similar analysis for the model LV2, when the LV effects
are proportional to the neutrino energy squared.
The sensitivity regions for the cases where oscillations are due to LV effects only,
and where standard oscillations are modified by LV effects, are very similar in shape
to those shown in Figs. \ref{lvsim:fig:LVsenseLV1} and \ref{lvsim:fig:LVsensen=1}
respectively.
In particular, the experimental best-fit point shown by a triangle in Fig.
\ref{lvsim:fig:LVsensen=1} again lies within the sensitivity region when standard
oscillations are modified by LV effects.
In this case, the sensitivity region again extends to both $\Delta m^{2}=0$ and
$\Delta \eta =0$.
The upper bound on $\Delta \eta $ from this sensitivity region is given in Table \ref{lvsim:tab:diagbounds}.

\subsubsection{Model LV3}

\begin{figure}
\begin{center}
\includegraphics[width=6.8cm]{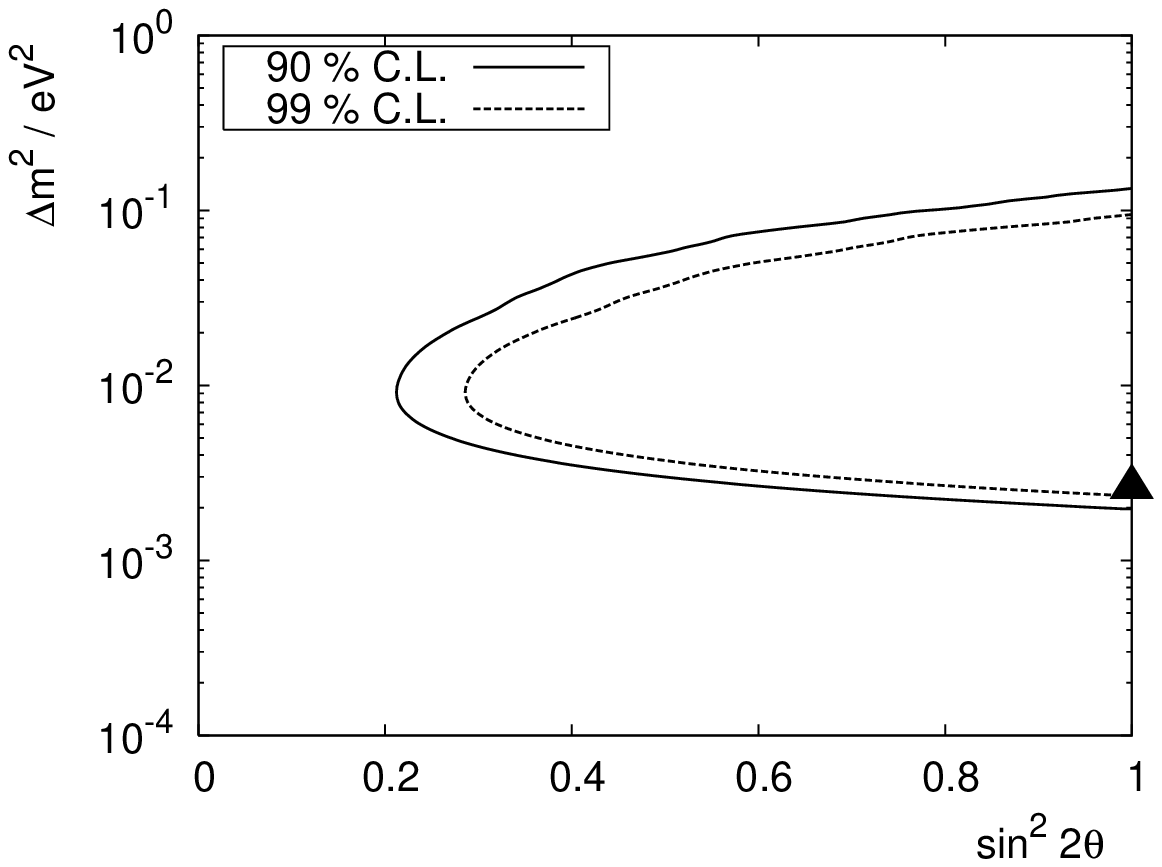}
\includegraphics[width=6.8cm]{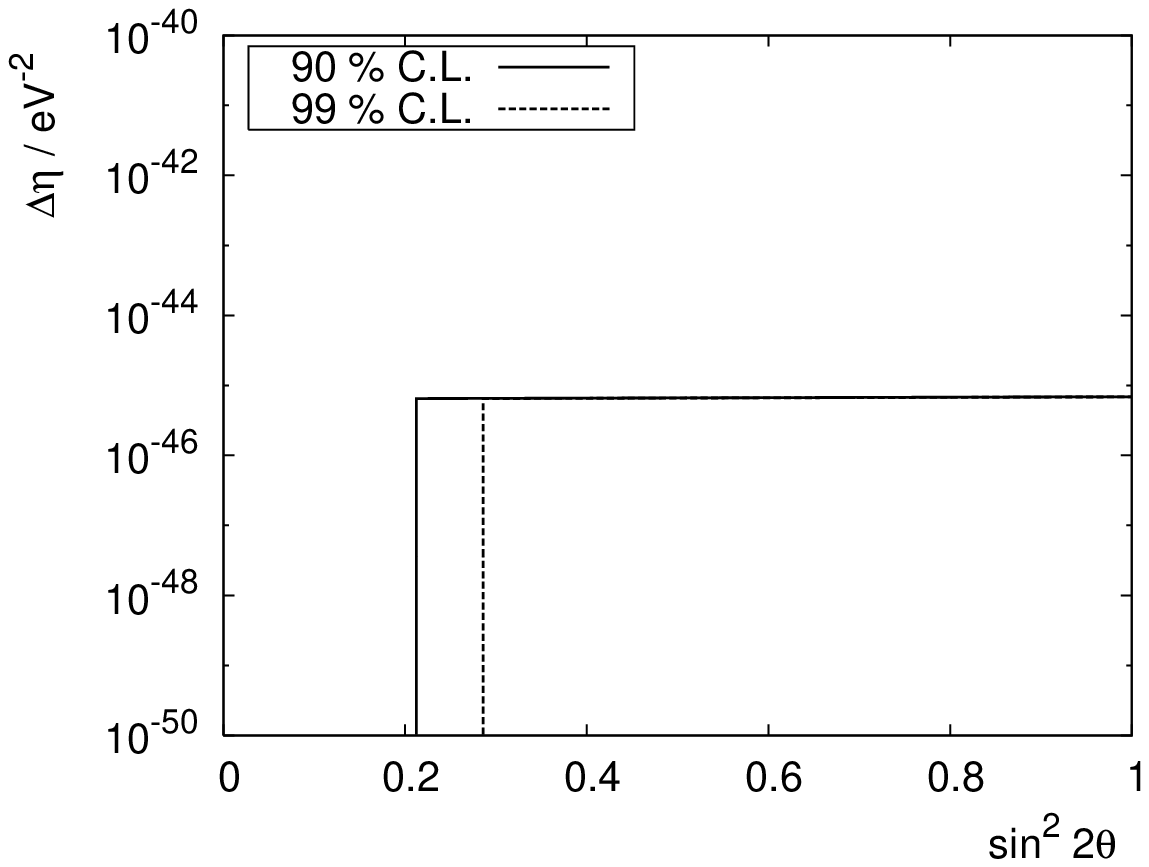}\vspace{2cm}
\includegraphics[width=6.8cm]{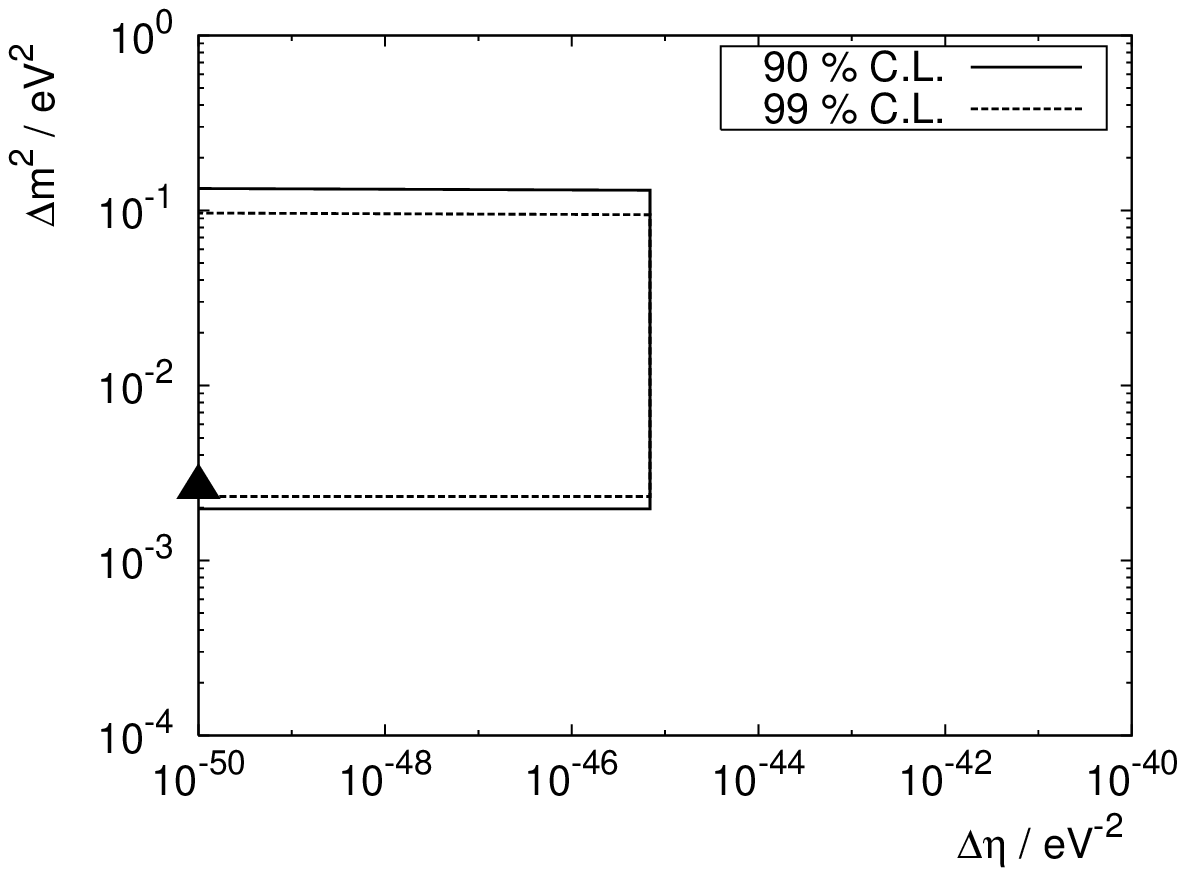}
\end{center}
\caption{Sensitivity contours for model
LV3 at 90 and 99 percent confidence level for standard
oscillations plus LV effects proportional to the neutrino
energy cubed.
The triangle denotes the best-fit value
of $\Delta m^{2}$ with no LV effects \cite{Gonzalez-Garcia:2003qf}.}
\label{lvsim:fig:LVsensen=3}
\end{figure}

The final model we consider has $n=3$. If we set $\Delta m^{2}=0$,
so that oscillations are due to LV effects only,
then we find that we are unable to derive any meaningful sensitivity contours, so
we do not consider this possibility further.
For $\Delta m^{2}\neq 0$, the sensitivity contours are shown in Fig.
\ref{lvsim:fig:LVsensen=3}.
The sensitivity contours in this case have a rather simpler shape than in
models LV1 and LV2.
Again the experimental point of
best fit is included in the sensitivity region but in this case,
$\Delta m^{2}=0$ is not included, so it seems that we are
not sensitive to the case when oscillations arise from LV effects only for
this energy dependence. This is in direct agreement with the first part of our
analysis as we were unable to find any meaningful results for this
model when we considered LV effects only. We note however that
$\Delta \eta =0$ is contained in the sensitivity region.
The upper bound on $\Delta \eta $ in this sensitivity region is given in Table \ref{lvsim:tab:diagbounds}.

\section{Discussion and conclusions}
\label{sec:conc}

We have shown that neutrino telescopes, such as ANTARES, will be able to place
stringent bounds on LV parameters which modify the dispersion relation for
massive neutrinos.
Table \ref{lvsim:tab:diagbounds} shows
the upper bounds for this model for the various energy dependences we have studied. We
found that if the LV effects were proportional to either the
neutrino energy or the neutrino energy squared, then
we are sensitive to
oscillations induced from these effects only, with no contributions from
standard oscillations. However, for the case of $n=3$, then we
cannot measure the effects of LV only in atmospheric neutrino oscillations.
For all values of $n$ studied, we have been able to produce sensitivity
curves for standard atmospheric neutrino oscillations modified by subdominant
LV effects.
We should stress that although our simulations are based on the ANTARES neutrino
telescope, we expect that other neutrino telescopes, such as AMANDA and ICECUBE,
will also be able to probe these models.

From a naturalness perspective, we would expect
any LV effects to be suppressed by the Planck mass, $10^{19}$ GeV.
Since our LV parameters $\Delta \eta $ contain powers of the Planck mass already,
this gives
$\Delta\eta\sim10^{-28(n-1)}$ eV$^{1-n}$ for $n\geq2$.
These values are also given in table \ref{lvsim:tab:diagbounds}
for comparison.
We note that for the
case $n=2$, the bounds obtained from our simulations surpass the
expected value of the LV parameter by seven orders of magnitude
and so ANTARES may be able to observe these effects or rule out
this model. For the case $n=3$, however, the bound from our
simulations is nine orders of magnitude larger than  expected
and so we cannot rule out this model using atmospheric neutrinos.
In this case better sensitivity can be obtained using higher-energy,
astrophysical neutrinos \cite{Hooper:2005jp}.
The situation is somewhat more complicated if $n=1$ as this
corresponds to $\alpha=0$ in (\ref{lvtheory:eqn:parmoddisp}) and
so we have no indication as to the value of $\Delta\eta$ except to
say that it should be small.
In all three cases, we can see from table \ref{lvsim:tab:diagbounds}
that our estimates based on the values of the LV parameters at which LV
effects become comparable to standard neutrino oscillations are very similar
to the upper bounds we find from our sensitivity curves.

Our results for the case $n=1$ are comparable with the model
investigated in \cite{Fogli:1999fs,Gonzalez-Garcia:2004wg} with
Super-Kamiokande and K2K data.
We find that we are not able to significantly improve on their results; however
an analysis of atmospheric neutrino data from ANTARES should be able to confirm their upper
bounds.
Although in a rather different LV model, our results are also
comparable in magnitude to those found by the MACRO collaboration \cite{Giorgini:2005zd}.
In that paper, the argument of the sine term in the oscillation probability is identical
to that considered here and they found an upper bound on
$\Delta \eta$ in the range $10^{-24}~<~\Delta
\eta~<~10^{-26}$, similar to that found here and shown in
table \ref{lvsim:tab:diagbounds}. Our results are also comparable to those found in \cite{kelley:2006}
using data from the AMANDA experiment.
There, an upper bound of $2.1 \times 10^{-27}$ was placed on the LV parameter.

In \cite{morgan:2004vv}, we studied in depth the spectra of numbers of events for
standard atmospheric neutrino oscillations with quantum decoherence.
Those spectra had some of the same key features as the ones presented here in Figs.
\ref{lvsim:fig:LVspectraLVn=1}-\ref{lvsim:fig:LVspectran=2}, in particular,
a suppression in the expected number of atmospheric neutrino events observed by ANTARES.
The shape of the spectra do have some differences, but it remains to be seen whether they
will really be distinguishable experimentally.
At this point, in the absence of real data, one can only speculate on what might be observed,
but one can foresee two possibilities.
Firstly, it may be that ANTARES (or any other neutrino telescope)
does not observe any distinguishable difference in numbers of
events compared with the standard atmospheric neutrino oscillation model.
We argued in \cite{morgan:2004vv} that such a null observation would be able to place
very stringent bounds on quantum decoherence effects, and we draw the same conclusion here
from our event spectra.
Secondly, it may be the case that ANTARES or an equivalent neutrino telescope does indeed
find evidence for a suppression in the number of events compared with the standard picture.
In this situation it is likely to be very challenging experimentally to disentangle the possible
source of new physics.

\section*{Acknowledgements}
The authors acknowledge the ANTARES Collaboration for
supplying a parametrization of the response of a neutrino telescope.
This work was supported by PPARC, and DM was supported by a PhD studentship from the University of Sheffield.
We would like to thank Nick Mavromatos for many helpful discussions.


\begin{thebibliography}{10}

\bibitem{Kiefer:2005uk}
C.~Kiefer,
{\it {Ann. Phys. }} {\bf {15}}, 129 (2005), {\tt {gr-qc/0508120}}.

\bibitem{Amelino-Camelia:2004hm}
G.~Amelino-Camelia,
{\it {Lect. Notes Phys. }}  {\bf {669}}, 59 (2005), {\tt {gr-qc/0412136}}.

\bibitem{Mavromatos:2004sz}
N.~E.~Mavromatos,
{\it {Lect. Notes Phys. }} {\bf {669}}, 245 (2005), {\tt {gr-qc/0407005}}.

\bibitem{morgan:2004vv}
D.~Morgan, E.~Winstanley, J.~Brunner, and L.~F. Thompson,
{\it {Astropart. Phys. }} {\bf {25}}, 311 (2006), {\tt {astro-ph/0412618}}.

\bibitem{Hooper:2004xr}
D.~Hooper, D.~Morgan, and E.~Winstanley,
{\it {Phys. Lett. }} {\bf {B609}}, 206 (2005), {\tt {hep-ph/0410094}}.

\bibitem{Hooper:2005jp}
D.~Hooper, D.~Morgan, and E.~Winstanley,
{\it {Phys. Rev. }} {\bf {D72}}, 065009 (2005), {\tt {hep-ph/0506091}}.

%
%
%
%
%


\bibitem{Fogli:1999fs}
G.~L.~Fogli, E.~Lisi, A.~Marrone and G.~Scioscia,
{\it {Phys. Rev. }} {\bf {D60}}, 053006 (1999), {\tt {hep-ph/9904248}}.

\bibitem{Gonzalez-Garcia:2004wg}
M.~C.~Gonzalez-Garcia and M.~Maltoni,
{\it {Phys. Rev. }} {\bf {D70}}, 033010 (2004), {\tt {hep-ph/0404085}}.

\bibitem{Giorgini:2005zd}
G.~Battistoni et al., {\it {Phys. Lett. }} {\bf {B615}}, 14 (2005),
{\tt {hep-ex/0503015}};
\newline
M.~Giorgini, {\it {Czech. J. Phys.}} {\bf{56}} A291 (2006), {\tt {hep-ex/0512075}}.

\bibitem{Korolkova:2004pg}
E.~V. Korolkova (ANTARES Collaboration),
{\it {Nucl. Phys. Proc. Suppl. }} {\bf {136}}, 69 (2004), {\tt {astro-ph/0408239}};
\newline
M.~Spurio (ANTARES Collaboration),
{\it{Status report of the ANTARES project}},
{\tt{hep-ph/0611032}}.

\bibitem{Ahrens:2002dv}
J.~Ahrens et al. (ICECUBE Collaboration),
{\it {Nucl. Phys. Proc. Suppl. }} {\bf {118}}, 388 (2003), {\tt {astro-ph/0209556}};
\newline
A.~Goldschmidt (ICECUBE Collaboration),
{\it {Nucl. Phys. Proc. Suppl. }} {\bf {110}}, 516 (2002);
\newline
P.~Desiati (ICECUBE Collaboration),
{\it{IceCube: Toward a km$^{3}$ Neutrino Telescope}},
{\tt{astro-ph/0611603}}.

\bibitem{Resvanis:1993gx}
L.~K. Resvanis (NESTOR Collaboration),
{\it {Nucl. Phys. Proc. Suppl. }} {\bf {35}}, 294 (1994).

\bibitem{Sapienza:2006en}
P.~Sapienza (NEMO Collaboration),
{\it{Status of the NEMO project}},
{\tt{astro-ph/0611105}}.

\bibitem{Wheeler:1}
J.~A. Wheeler,
{\it {Superspace and the nature of quantum geometrodynamics}},
in: C. DeWitt, J. A. Wheeler (Eds.), Battelle Rencontres: 1967
  Lectures in Mathematical Physics  (W. A. Benjamin, New York, 1968).

\bibitem{Horowitz:2004rn}
G.~T. Horowitz,
{\it {New J. Phys. }} {\bf {7}}, 201 (2005), {\tt {gr-qc/0410049}}.

\bibitem{Mohaupt:2002py}
T.~Mohaupt,
{\it {Lect. Notes Phys. }} {\bf {631}}, 173 (2003), {\tt {hep-th/0207249}};
\newline
J.~H. Schwarz,
{\it {Introduction to superstring theory}}
in: Proceedings of Techniques and Concepts of High-Energy Physics,
  {U.S.A.}, edited by H. B. Prosper and M. Danilov (NATO, 2000),
  {\tt {hep-ex/0008017}}.

\bibitem{Dhar:2001a}
A.~Dhar,
{\it {Current Science}} {\bf {81}}, 1598 (2001);
\newline
E.~D'Hoker,
{\it {Mod. Phys. Lett. }} {\bf {A6}}, 745 (1991);
\newline
E.~D'Hoker and P.~S. Kurzepa,
{\it {Mod. Phys. Lett.}} {\bf {A5}}, 1411 (1990);
\newline
J.~Distler and H.~Kawai,
{\it {Nucl. Phys.}} {\bf {B321}}, 509 (1989);
\newline
J.~R. Ellis, N.~E. Mavromatos, and D.~V. Nanopoulos,
{\it {Mod. Phys. Lett. }} {\bf {A12}}, 1759 (1997), {\tt {hep-th/9704169}};
\newline
N.~E. Mavromatos and J.~L. Miramontes,
{\it {Mod. Phys. Lett. }} {\bf {A4}}, 1847 (1989).

\bibitem{Rovelli:1997yv}
C.~Rovelli,
{\it {Living Rev. Rel. }} {\bf {1}}, 1 (1998), {\tt {gr-qc/9710008}}.

\bibitem{Greenberg:2002uu}
O.~W. Greenberg,
{\it {Phys. Rev. Lett. }} {\bf {89}}, 231602 (2002), {\tt {hep-ph/0201258}}.

\bibitem{Rovelli:1994ge}
C.~Rovelli and L.~Smolin,
{\it {Nucl. Phys. }} {\bf {B442}}, 593 (1995), {\tt {gr-qc/9411005}}.

\bibitem{Mattingly:2005re}
D.~Mattingly,
{\it {Living Rev. Rel. }} {\bf {8}}, 5 (2005), {\tt {gr-qc/0502097}}.

\bibitem{Coleman:1998ti}
S.~R.~Coleman and S.~L.~Glashow,
{\it {Phys. Rev. }} {\bf {D59}},  116008 (1999), {\tt {hep-ph/9812418}}.

\bibitem{Gambini:1998it}
J.~R. Ellis, N.~E. Mavromatos, and D.~V. Nanopoulos,
{\it {Gen. Rel. Grav. }} {\bf {32}}, 127 (2000), {\tt {gr-qc/9904068}};
\newline
R.~Gambini and J.~Pullin,
{\it {Phys. Rev. }} {\bf {D59}}, 124021 (1999), {\tt {gr-qc/9809038}}.

\bibitem{Ellis:1999yd}
J.~Albert  et al.  (MAGIC Collaboration),
{\it {Probing quantum gravity using photons from a Mkn 501 flare observed by MAGIC}},
{\tt {arXiv:0708.2889 [astro-ph]}};
\newline
J.~Bolmont, J.~L.~Atteia, A.~Jacholkowska, F.~Piron and G.~Pizzichini,
{\it {Study of time lags in HETE-2 gamma-ray bursts with redshift: search for
astrophysical effects and quantum gravity signature}},
{\tt {astro-ph/0603725}};
\newline
J.~R. Ellis, N.~E. Mavromatos, and D.~V. Nanopoulos,
{\it {Probing models of quantum space-time foam}},
in: Proceedings of beyond the desert 1999: Accelerator,
non-accelerator and space approaches into the next millenium, edited by H. V.
Klapdor-Kleingrothaus and I. V. Kirvosheina (IOP, 2000),
{\tt {gr-qc/9909085}};
\newline
J.~Ellis, N.~E.~Mavromatos, D.~V.~Nanopoulos, and A.~S.~Sakharov,
{\it {Astron. Astrophys. }} {\bf {402}} 409 (2003), {\tt {astro-ph/0210124}};
\newline
J.~Ellis, N.~E.~Mavromatos, D.~V.~Nanopoulos, A.~S.~Sakharov and E.~K.~G.~Sarkisyan,
{\it {Astropart. Phys.}} {\bf {25}} 402 (2006), {\tt {astro-ph/0510172}};
\newline
R.~Lamon, N.~Produit and F.~Steiner,
{\it {Study of quantum gravity effects in INTEGRAL gamma-ray bursts}},
{\tt {arXiv:0706.4039 [gr-qc]}}.

\bibitem{Lambiase:2003bq}
G.~Lambiase,
{\it {Mod. Phys. Lett. }} {\bf {A18}}, 23 (2003), {\tt {gr-qc/0301058}}.

\bibitem{Christian:2004xb}
J.~Christian,
{\it {Phys. Rev. }} {\bf {D71}}, 024012 (2005), {\tt {gr-qc/0409077}}.

\bibitem{kelley:2006}
J.~L.~Kelley (ICECUBE Collaboration),
{\it{Testing Lorentz invariance using atmospheric neutrinos and AMANDA-II}},
in: Proceedings of the first workshop on exotic physics with neutrino telescopes,
edited by C. de los Heros (Uppsala University, 2007),
{\tt{astro-ph/0701333}}.

%
%
%
%
%
%
%
%
%
%
%
%

\bibitem{Gonzalez-Garcia:2006mh}
M.~C.~Gonzalez-Garcia,
{\it{Non-standard neutrino oscillations at IceCube}},
in: Proceedings of the first workshop on exotic physics with neutrino telescopes,
edited by C. de los Heros (Uppsala University, 2007),
{\tt{astro-ph/0701333}};
\newline
M.~C.~Gonzalez-Garcia, F.~Halzen and M.~Maltoni,
{\it {Phys. Rev. }} {\bf {D71}}, 093010 (2005),
{\tt{hep-ph/0502223}}.

\bibitem{Blennow:2007pu}
M.~Blennow, T.~Ohlsson and J.~Skrotzki,
{\it {Phys. Lett.}} {\bf {B660}}, 522 (2008),
{\tt{hep-ph/0702059}}.

\bibitem{Blennow:2005qj}
M.~Blennow, T.~Ohlsson and W.~Winter,
{\it{Eur. Phys. J.}} {\bf{C49}}, 1023 (2007),
{\tt{hep-ph/0508175}}.

\bibitem{Amelino-Camelia:2002fw}
G.~Amelino-Camelia, {\it {On the fate of Lorentz symmetry in loop quantum gravity and
noncommutative spacetimes}}, {\tt {gr-qc/0205125}}.

\bibitem{Amelino-Camelia:2003ex}
G.~Amelino-Camelia, J.~Kowalski-Glikman, G.~Mandanici, and A.~Procaccini,
{\it {Int. J. Mod. Phys. }} {\bf {A20}}, 6007 (2005), {\tt {gr-qc/0312124}}.

\bibitem{Ashie:2004mr}
Y.~Ashie et al. (Super-Kamiokande Collaboration),
{\it {Phys. Rev. Lett. }} {\bf {93}}, 101801 (2004), {\tt {hep-ex/0404034}}.

\bibitem{Gonzalez-Garcia:2003qf}
M.~C. Gonzalez-Garcia and C.~Pena-Garay,
{\it {Phys. Rev. }} {\bf {D68}}, 093003 (2003), {\tt {hep-ph/0306001}}.

%
%
%
%
%

\bibitem{Brunner:1999ig}
F.~Blondeau and L.~Moscoso (for the ANTARES Collaboration),
{\it {Detection of atmospheric neutrino oscillations with a 0.1 km${}^{2}$ detector :
the case for ANTARES}},
NOW 98 conference, Amsterdam, Netherlands (1998);
\newline
J.~Brunner (for the ANTARES Collaboration),
{\it {Measurement of neutrino oscillations with neutrino telescopes}},
15th International Conference on Particles and Nuclei (PANIC 99), Uppsala, Sweden (1999);
\newline
C.~Carloganu (for the ANTARES Collaboration),
{\it {Nucl. Phys. Proc. Suppl.}} {\bf {100}} 145 (2001).

\bibitem{Gaisser:2001jw}
T.~K. Gaisser, T.~Stanev, M.~Honda, and P.~Lipari,
{\it {Primary spectrum to 1-TeV and beyond}},
in: Proceedings of 27th International Cosmic Ray Conference
  (ICRC2001), Hamburg, Germany (2001).

\bibitem{Araki:2004mb}
T.~Araki et al. (KamLAND Collaboration),
{\it {Phys. Rev. Lett. }} {\bf {94}}, 081801 (2005), {\tt {hep-ex/0406035}}.

\bibitem{Aliu:2004sq}
E.~Aliu et al. (K2K Collaboration),
{\it {Phys. Rev. Lett. }} {\bf {94}}, 081802 (2005), {\tt {hep-ex/0411038}}.


\end{thebibliography}
\end{document}